\documentclass{aa}
\usepackage{amsmath}
\usepackage{amssymb}
\usepackage{natbib}
\usepackage{xcolor}
\usepackage{import}
\usepackage[varg]{txfonts}

\usepackage{graphicx}
\usepackage[caption=false]{subfig}
\usepackage{float}
\usepackage{amsfonts}

\def\apjs{{\rm ApJS}}

\def\simlt{\lower.5ex\hbox{$\; \buildrel < \over \sim \;$}}
\def\simgt{\lower.5ex\hbox{$\; \buildrel > \over \sim \;$}}
\def\simpropto{\lower.5ex\hbox{$\; \buildrel \propto \over \sim \;$}}

\newcommand{\bfx}{{\pmb{x}}}

\newcommand{\bfr}{{\pmb{r}}}

\newcommand{\bx}{\pmb{x}}

\newcommand{\btheta}{{\pmb{\theta}}}
\newcommand{\balpha}{{\pmb{\alpha}}}
\newcommand{\bgamma}{{\pmb{\gamma}}}

\newcommand{\grad}{\pmb{\nabla}}

\newcommand{\bell}{{\pmb{\ell}}}
\newcommand{\tr}{{\rm tr}}
\newcommand{\lya}{Ly$\alpha$}
\newcommand{\lyb}{Ly$\beta$}

\newcommand{\mat}[1]{\pmb{\mathrm{#1}}}

\newcommand{\Ave}[1]{{\left\langle #1 \right\rangle}}

\def\aap{Astronomy \& Astrophysics}

\usepackage[normalem]{ulem}
\newcommand{\nt}[2][]{\def\old{#1}\def\new{#2}{\color{purple}\ifx\old\empty\else\sout{#1}\ifx\new\empty\else\ \fi\fi #2}}

\begin{document}

\title{Reconstructing the Gravitational Lensing Potential from the Lyman-$\alpha$ Forest }
\author{
R. Benton Metcalf \inst{1,2}\thanks{E-mail: robertbenton.metcalf@unibo.it} 
\and 
Nicolas Tessore\inst{3}
\and
Rupert A.C. Croft \inst{4}
}

\institute{
Dipartimento di Fisica \& Astronomia, Universit\`a di Bologna, via Gobetti 93/2, 40129 Bologna, Italy 
\and
INAF-Osservatorio Astronomico di Bologna, via Ranzani 1, 40127 Bologna, Italy 
\and
Jodrell Bank Centre for Astrophysics,
University of Manchester,
Alan Turing Building,
Oxford Road,
Manchester, M13~9PL, UK
\and
McWilliams Center for Cosmology, Department of Physics, 
Carnegie Mellon University, Pittsburgh, PA 15213, USA
}

\date{Accepted . Received ; in original form }

\abstract{
We demonstrate a method for reconstructing the weak lensing potential from the Lyman-$\alpha$ forest data.  
We derive an optimal estimator for the lensing potential on the sky based on the correlation between pixels in real  space.  This method effectively deals with irregularly spaced data, holes in the survey, missing data and inhomogeneous noise.
We demonstrate an implementation of the method with simulated spectra and weak lensing.   It is shown that with a source density of $\simgt 0.5$ per square arcminutes and $\sim 200$ pixels in each spectrum ($\lambda / \Delta\lambda = 1300$) the lensing potential can be reconstructed with high fidelity if the relative absorption in the spectral pixels is signal dominated.   When noise dominates the measurement of the absorption in each pixel the noise in the lensing potential is higher, but for reasonable numbers of sources and noise levels and a high fidelity map the lensing potential is obtainable.  The lensing estimator could also be applied to lensing of the Cosmic Microwave Background (CMB), 21 cm intensity mapping (IM) or any case in which the correlation function of the source can be accurately estimated.  
}


\maketitle


\section{Introduction}
\label{sec:intro}

Weak gravitational lensing \citep{bart01,bart17} describes the distortion of
background sources of light by foreground matter in the regime where
statistical averaging is needed to detect the effect. As any object at
cosmological distances can be lensed, a variety of background
sources apart from
galaxies \citep{valdes83,kaiser92,troxel18}, such as radio sources
\citep{chang04,harrison16}, or even already-lensed images themselves
\citep{birrer18} can be studied with weak lensing techniques.
These examples are discrete objects and statistical measures
of weak lensing \citep[e.g.,][]{miralda91},
that are applied to those observations
treat them as such. This includes estimators of the foreground
lensing matter distribution \citep[e.g.,][]{kaiser93,niall20}.
On the other hand, a continuous
background source field at cosmological distances can also be lensed,
and a whole class of estimators has been developed to deal with this
case, starting with the Cosmic Microwave Background (CMB) as
the source field \citep[e.g.,][]{bernard97,1997ApJ...489....1M,1998ApJ...492L...1M,zald99,huok02,schaan19}.
In the continuous case,
it is the statistical properties of the field which change as it is
lensed, with the both the scale and isotropy of correlations imparting
information on the foreground matter
\citep[see reviews by][]{lewis06,hanson10}. In the present paper
we develop and test the first estimator appropriate to a particular cosmological
field for which lensing has not yet been observed, the Lyman-$\alpha$ (\lya) forest.

After the CMB, another continuous
field which has been studied in this context is the emission from neutral
hydrogen in the Universe close to the epoch of reionization,
investigated using the 21 cm radio line \citep{madau97,furl06}. In this case
the field is in three dimensions as the use of a discrete line yields
redshift information, and datacubes of line intensity are known as
intensity maps \citep{wyithe07}. Even at lower redshifts where
galaxies act as more discrete
sources of 21cm emission, intensity maps can be made without identifying
individual sources, and trace out large-scale structure more efficiently
\citep{chang10,anderson18}.
Weak lensing using estimators appropriate to 21cm intensity maps
have been developed \citep[][see also \citealp{pour14} for the low redshift case]{zz06,metcalf07,2015MNRAS.448.2368P,2018JCAP...07..046F}.
They are generalizations of the
estimators applied to the CMB, and as in that case act on a continuous
lensed source field, usually employing Fourier techniques. 

The neutral hydrogen density field also absorbs radiation,
and the \lya\ line can be used to observe it in the spectra of background
galaxies and quasars (which we call ``sources'' here). The absorption
signatures are the Lyman-$\alpha$ forest \citep{rauch98,prochaska19}, and are good
tracers of the overall cosmological density field
on scales larger than the gas pressure
smoothing scale (of order 0.1 Mpc, \citealp{peeples10}). Clustering of the
\lya\ forest  \citep[see e.g.][for some early work]{croft98,hui99,mcdonald99} 
is now a mainstream cosmological probe, for example
having been used to measure the Baryon Oscillation scale at $z=2$-$3$
\citep{busca13,slosar13,ebosslyabao},
the power spectrum of mass fluctuations \citep[e.g.,][]{chabanier19}, and
constrain the properties of dark matter \citep[e.g.,][]{irsic17}).  By including high redshift galaxies as well as quasars a higher source density can be reached at the expense of more telescope time.  \citet{clamato18} and \citet{2020arXiv200210676N} have used this higher density to construct 3 dimensional maps of the neutral atomic hydrogen (HI).

The \lya\ forest is well understood, and can be simulated in the context
of CDM models \citep[e.g.,][]{bolton17,rogers18,chab20}.
\citet{2011MNRAS.415.2257M,2015MNRAS.448.1403M} have studied
the problem of measuring the power spectrum from the \lya\ forest.
Many significant observational
datasets either exist \citep[][]{ebossdr16,clamato18,2020arXiv200210676N}, are planned
or are in progress
\citep{desi16, dalton18, mse18}
The forest is also the first weak lensing probe for which data exists that
has full spectroscopic information, thus bypassing the difficulties
of galaxy lensing associated with photometric redshifts \citep{bordoloi10,
cunha14}.
As such it represents a useful source field for weak lensing, and
\cite{croft18} and \cite{metcalf18} explored what will be possible, but
without developing
an estimator appropriate for the specific geometry of the forest.

The \lya\ forest requires development of a new lensing estimator because
although it samples a continuous three-dimensional field, the sampling
is not uniform or space-filling.
The absorption is imprinted in the spectra of sources
(galaxies or quasars), which have an angular extent which is much smaller
than the separation between them. The angular coverage is therefore
effectively a set of delta functions, although the spectra are continuous
along the line of sight (excepting masked contaminated regions).  There is also noise in the pixel values in these
spectra, including CCD readout noise, Poisson photon noise,
flux calibration errors and sky subtraction errors
\citep[see e.g.,][]{bautista15}.
and also potentially large-scale correlated
noise  from uncertainty in the zero level
\citep[continuum fitting,][]{blomqvist15}. The angular positions of the
sources are usually determined
to a much smaller uncertainty  \citep[$<0.05$ arcsec,][]{ahn12} than the
mean lensing deflection angle of a few arcmins \citep{lewis06}
however, and are not a source
of noise in themselves. 

We are therefore not able to apply the usual quadratic estimators \citep[e.g.,][]{huok02,metcalf&white2007}
to forest lensing, as these rely on uniform coverage with a regular
grid. In our previous work reconstructing the lensing potential
from simulated \lya\ forest sightlines \citep{croft18}, we worked with a toy
model mock survey,  with all sources  on an evenly
spaced grid, a situation not applicable to observational data. In
practice, the angular distribution of quasar and galaxy sources is not only
irregular, but can be extremely sparse, with the data sets with
largest current total numbers having a mean separation between sources
of order 10 arcmin \citep{blom19}. In general
it would be extremely useful to have
an estimator which can deal with a variety of different survey geometries
and sampling, and that is our aim in this paper.

The method we develop
is general, and although we focus here on the \lya\ forest,
it has applications to the weak lensing of
other sources that can be treated as continuous and isotropic such as
the CMB \citep{lewis06} or spectral line intensity mapping 
\citep[SLIM, e.g.,][]{kovetz17}.
In fact the same method can be applied to any linear distortion in the 3D correlation function relative to expectations, including redshift distortions, or the scale of the Baryon Acoustic
Oscillations (BAO).

The outline of our paper is as follows.
In the next section we review the necessary basics of \lya\ observations.   In section~\ref{sec:lensing} we calculate the effect lensing has on the second order correlations between 
the \lya\ absorption in the pixels in the spectra of quasars and galaxies.  We derive our quadratic estimator for the lensing potential and discuss some of its properties in section~\ref{sec:QE}.   The methods used to simulate data are described in section~\ref{sec:simulation_method} and in the appendices.   In section~\ref{sec:demonstrations}
the estimator is tested on various simulated data sets.  A final discussion is in section~\ref{sec:discussion}.  The appendices contain details about simulating data and on using different expansions of the lensing potential in the estimator.

\section{The \lya\ forest}
\label{sec:lya_forest}

Neutral hydrogen will absorb photons through the \lya\ transition ($\lambda_\alpha = 1215.67$\,\AA ) along the sight line to high redshift sources.  Because of cosmological redshifting this absorption will appear at an observed wavelength, $(1+z)\lambda_\alpha$, where $z$ is below the redshift of the source.  In general the redshift range of the \lya\ forest is limited to $\simgt 2$ because of atmospheric opacity at lower wavelengths.  In individual cases the redshift range is limited on the high side by the redshift of the source and on the low side 
by confusion with absorption by the \lyb\ line if not by the atmosphere.  Damped \lya\ systems cannot be used so they need to be masked out \citep[see e.g.][]{font2012}.  As a result of these considerations the volume probed by the \lya\ forest is very irregularly sampled.

The flux at observed wavelength or pixel $\lambda$ in the spectra of the $q$th source can be written \citep{2013A&A...552A..96B,2011JCAP...09..001S,2013JCAP...04..026S}
\begin{equation} \label{eq:absorption}
f_{q,\lambda} = C_q(\lambda) \, \bar{F}(\lambda) \, (1 + \delta_{q\lambda} )  + n_{q,\lambda} \;,
\end{equation}
where $ C_q(\lambda)$ is the unabsorbed spectrum of $q$th source, $\bar{F}(\lambda)$ is the mean transmission in the forest at $\lambda$ and $n_{q,\lambda}$ is uncorrelated pixel noise.  Other factors such as the sky background, spectrograph transmissivity and PSF are either negligible or can be absorbed into these quantities.  The quantities $\delta_{q\lambda}$ are the fluctuations in the absorption which to first order are proportional to fluctuations in the column density of HI.  For the purposes of this paper however, the details of the relationship between  $\delta_{q\lambda}$ and the HI density or the dark matter density are not important.  It will only be assumed that the statistical distribution of $\delta_{q\lambda}$ is isotropic and that its correlation function can be measured or modeled.

A two parameter model for the unabsorbed spectra, $C_q(\lambda)$, the $\delta_{q\lambda}$'s and their covariance matrix can be found using different methods, for example those of \citet{2013A&A...552A..96B} and \citet{2011JCAP...09..001S,2013JCAP...04..026S}, or by fitting template spectra in the case of sources that are galaxies \citep{2020arXiv200210676N,clamato18}.  One method is to find the maximum likelihood solution for $\delta_{q\lambda}$ by brute force minimization and then approximate the covariance matrix as the Fisher matrix found numerically at the maximum likelihood solution.  $\bar{F}(\lambda)$ is found by averaging across all sources in the data or perhaps from another survey with
higher signal to noise and higher resolution spectra \citep[e.g.,][]{faucher08}.  The fitting of these quantities will induce covariance between pixels that would not otherwise be there solely due to observational noise.

In this paper we will assume that the modeling of $C_q(\lambda)$ and $\bar{F}(\lambda)$ has already been done to find $\delta_{q\lambda}$ and the covariance matrix ${\bf N}$.  It may also be advantageous to transform the data using a function that makes the distribution of $\delta_{q\lambda}$ values strictly
Gaussian \citep[e.g.,][]{croft98}, but we do not investigate that here.

\section{The effect of lensing on the \lya\ forest}
\label{sec:lensing}

As light passes through structures in the Universe it will be deflected from a geodesic of the unperturbed, homogeneous metric.  One way of representing this is to imagine many planes that are perpendicular to the unperturbed light ray.   At each plane the ray is deflected.  In the Born approximation the perpendicular position at redshift $z$ can be written as
\begin{align}
 \pmb{x}_\perp (z) & = \left[ \pmb{\theta} D(z) - \sum_{m < z} D(z_m,z) ~  \hat{\balpha}_m(\pmb{\theta}) \right] \\
& = \left[ \pmb{\theta} - \sum_{m < z}  \balpha_m(\pmb{\theta}) \right] D(z) 
\end{align}
where $\hat{\balpha}_m(\pmb{\theta})$ is the deflection at the $m$th plane  and the sum is over all planes at lower redshift \citep{2014MNRAS.445.1954P,SEF92}.  The  angular position of the image is $\btheta$, the angular diameter distance between redshifts $z_1$ and $z_2$ is $D(z_1,z_2)$, and $D(z) = D(0,z)$.   
The usual scaled deflection angle~$\balpha_m(\pmb{\theta})$ is obtained by rescaling all deflection angles to redshift~$z$.
The total deflection for a light ray originating at redshift~$z$ is hence the sum over all lensing planes below~$z$,
\begin{align}
\balpha(\pmb{\theta},z) \equiv \sum_{m < z}  \balpha_m(\pmb{\theta}) .
\end{align}
In addition for this paper we will make the approximation that all  the deflections of significance take place at a redshift smaller than the redshift of the \lya\ absorption that we will consider.  This is a good first approximation which can be relaxed when the data warrants it.

In the weak lensing limit, the deflection field is curl free so a {\it lensing potential} can be defined, the gradient of which is the deflection
\begin{align}
\balpha(\btheta,z) = \grad \phi (\btheta,z).
\end{align}
The gradient $\grad$ here is taken with respect to the angular coordinates~$\btheta$.
The potential is related to the convergence, $\kappa(\btheta)$, by a Poisson equation
\begin{equation}
\nabla^2 \phi(\btheta,z) = 2\kappa(\btheta,z).
\end{equation}
The convergence is the trace of the magnification matrix and in the weak lensing context can be interpreted as a weighted surface density on the sky,
\begin{align}
\kappa(\btheta) = \frac{3}{2} \frac{\Omega_m H_o}{c^2} \int_o^{\chi_s} d\chi ~ \left[ \frac{d_A(\chi)d_A(\chi,\chi_s) }{d_A(\chi_s)} \right] ~ \delta(\btheta,\chi)
\end{align}
where $\delta(\btheta,\chi)$ is the density contrast at the radial coordinate, comoving distance $\chi$, $d_A(\chi)$ is the comoving angular size distance,  $\Omega_m$ is the matter density parameter and $H_o$ is the Hubble parameter.



Gravitational lensing moves the apparent position of a source on the sky.  In the plane sky approximation (this approximation is relaxed in Appendix~\ref{app:sphereical}) the observed absorption $\delta_o$ at the radial position $r_\parallel$ and perpendicular position $r_\perp$ is, to first order,
\begin{equation}
\begin{split}
\delta_o(\pmb{x}_\parallel,\pmb{x}_\perp) &= \delta\left (\pmb{x}_\parallel,\pmb{x}_\perp - D_{s} \,\pmb{\alpha}(\pmb{\theta}) \right) \\  
 &= \delta(\pmb{x}) - D \, \balpha(\btheta) \cdot \nabla_\perp \delta(\pmb{x}) + \dots \\
 &= \delta(\pmb{x}) - D \, \grad\phi(\btheta) \cdot \grad_\perp \delta(\pmb{x}) + \dots
\end{split}
\end{equation}
where $\delta(\btheta)$ is the intrinsic absorption, as it would have appeared without lensing.
To first order in the deflection, the observed absorption correlation function will be changed by
\begin{align}
\langle \delta_o(\bfx) \delta_o(\bfx') \rangle  = &\, \langle \delta(\bx) \delta(\bx') \rangle - D \, \balpha(\bx) \cdot \grad_\perp \langle  \delta(\bfx) \delta(\bfx') \rangle  \nonumber \\
& -  D'\, \balpha(\bx') \cdot \grad_\perp' \langle  \delta(\bfx) \delta(\bfx')\rangle  \\
=&\, \xi(\bfr)  + \left[ D\, \balpha(\btheta) - D'\,\balpha(\btheta') \right] \cdot \grad_\perp \xi(r_\parallel,r_\perp) \\
= &\, \xi(\bfr)   + \left[ D\, \alpha^\sigma(\btheta) - D'\, \alpha^\sigma(\btheta') \right]  \left[ \frac{\partial r_\perp}{\partial\pmb{x}_\perp^\sigma} \frac{\partial\xi}{\partial r_\perp} \right] \\
\simeq &\, \xi(\bfr)   + \left[ d(z)\,  \balpha(\btheta) -  d(z')\, \balpha(\btheta') \right]  \cdot  \pmb{\gamma} \,\left[ \frac{\overline{D}^2}{r_\perp} \frac{\partial\xi}{\partial r_\perp} \right]  \\
\simeq &\, \xi(\bfr)   + \left[ d(z)\, \balpha(\btheta) - d(z')\, \balpha(\btheta') \right]  \cdot  \pmb{\gamma} \,\left[ \frac{1}{|\pmb{\gamma}|^2} \frac{\partial\xi}{\partial \ln r_\perp} \right]  \label{eq:corrfunction} \\
=&\, \xi (\bfr)  + \left[ d(z) \, \balpha(\btheta) - d(z')\, \balpha(\btheta') \right]  \cdot \bgamma \,\mathcal{G}(\bfr)  \label{eq:G} 
\end{align}
where $\bfr = (|\bfx_\parallel - \bfx_\parallel |,|\bfx_\perp - \bfx_\perp' |)$, $r =| \bfr |$,
$r_\perp = |\bfx_\perp - \bfx_\perp' |$, $\overline{D}$ is the average angular distance to the two redshifts  and $d(z) = D(z) / \overline{D}(z,z')$.  $\xi(r_\perp,r_\parallel)$ is the intrinsic, pre-lensing, correlation function.  Here $\bgamma$ is the angular separation of the two points $\bfx$ and $\bfx'$.   In deriving line (\ref{eq:corrfunction}) we have assumed that any two pixels that are strongly enough correlated to contribute to the lensing signal are at approximately the same angular size distance, $r_\parallel \ll D_s,D_s'$.  This approximation can be relaxed if necessary, but is probably a good one.
Line (\ref{eq:G}) serves to define the function $\mathcal{G}(D,r)$.   
We have written the intrinsic correlation function as a function of both $r_\parallel$ and $r_\perp$ to allow for redshift distortions caused by the radial velocities of the HI.

Our proposed method is based on modelling the deflection field~$\phi(\btheta)$ in a form that is linear in the coefficients~$\hat{\phi}_\ell$ for a set of basis functions~$f_\ell(\btheta)$,
\begin{equation}
    \phi(\btheta)
    = \sum_{\ell} \hat{\phi}_\ell \, f_\ell(\btheta) \;.
\end{equation}
The deflection field $\balpha = \grad\phi$ at angular position~$\btheta_i$ is then related to the model parameters $\hat{\phi}_j$ as
\begin{equation}\label{eq:alphaFT}
\alpha^\nu (\pmb{\theta}_i) = \frac{\partial}{\partial \theta_\nu} \phi(\pmb{\theta}_i) 
= \sum_{\ell} \mathcal{A}^\nu_{i\ell} \, \hat{\phi}_\ell \, ,
\end{equation}
where $\nu = 1, 2$ is the component of the deflection field, and $\mathcal{A}^{\nu}_{i\ell}$ is the gradient of the model's basis function~$f_\ell$ in the point~$\btheta_i$.
We will shortly show specific instances of these models for a number of well-known expansions of the potential.

Including the noise correlation matrix $N_{ij}$, the correlation~(\ref{eq:G}) between pixels $\delta_i$ and $\delta_j$ can be written
\begin{align}
\langle \delta_i \delta_j \rangle  &= \xi(\bfr_{ij}) + N_{ij}  + \sum_\ell  P^\ell_{ij}  \hat{\phi}_\ell \\
&= C_{ij}  + \sum_\ell  P^\ell_{ij}  \hat{\phi}_\ell \label{eq:ave_dd}
\end{align}
where
\begin{align} \label{eq:def_P}
P^\ell_{ij} = \sum_{\nu} \left( \mathcal{A}^\nu_{i\ell} - \mathcal{A}^\nu_{j\ell} \right) \gamma_\nu \, \mathcal{G}(D,\pmb{r}_{ij}).
\end{align}
and $\pmb{C}$ is the correlation matrix between pixels including intrinsic correlations, noise and correlations that come from the modeling procedure outlined in section~\ref{sec:lya_forest}.  
The $P^\ell_{ij}$ matrix will differ depending on the model for the potential and on how its derivatives are estimated.

For example, 
if the potential is modelled by a discrete Fourier transform (DFT) with coefficients $\tilde{\phi}_\bell$ and $\bell = (\ell_1, \ell_2)$, we can compute the deflection field,
\begin{equation}\label{eq:alphaDFT}
\balpha(\btheta)  = \vec{\grad} \phi(\btheta) 
=  \sum_{\bell} i \bell  \, e^{i \bell \cdot \btheta} \,\tilde{\phi}_\bell \;,
\end{equation}
and the $P^\bell_{ij}$ matrices~\eqref{eq:def_P},
\begin{equation} \label{eq:Pmatrix}
P^\bell_{ij} =  2 \left( \bell \cdot \btheta_{ij} \right) \sin\left( \frac{\bell \cdot \btheta_{ij}}{2}\right) e^{-i \bell \cdot \overline{\btheta}_{ij}} \mathcal{G}\left(D,\pmb{r}_{ij}\right) \;,
\end{equation}
where $\btheta_{ij} = \btheta_i - \btheta_j$ and $\overline{\btheta}_{ij} = (\btheta_i + \btheta_j)/2$. 

   More details for the implementation of this and  other expansions are given in Appendix~\ref{app:expansions}.   
 We use the Legendre expansion, section~\ref{app:legendre_expansion}, for our numerical demonstrations for reasons that will become clear later.  We have also implemented and tested a bilinear, finite difference expansion on a grid and a Chebyshev expansion, section~\ref{app:chebyshev_expansion}.  The advantage of the Legendre and Chebyshev expansions are that they do not impose any boundary conditions and, unlike the bilinear case, they have continuous derivatives so that the deflection field is continuous.

Note that two pixels $(i,j)$ in the spectrum of a single source will have the same angular position and so $P^\bell_{ij}$ will be zero.  Two pixels from different spectra will have uncorrelated noise, so $N_{ij}$ will be zero,
except perhaps through the estimate~\eqref{eq:absorption} of $\bar{F}(\lambda)$.  The intrinsic correlation between the absorption must be isotropic in angle, so $\xi(\pmb{r}_{ij})$ will be a function of only the absolute angular separation between the pixels and their radial separation.

\section{Quadratic Estimator}
\label{sec:QE}

We can express equation~(\ref{eq:ave_dd}) in matrix form 
\begin{equation} \label{eq:vector_ave}
\langle \pmb{\Delta} \rangle  = {\bf C}  + {\bf P}^\nu \tilde{\phi}_\nu
\end{equation}
with the definition
\begin{equation}
\left[ \pmb{\Delta} \right]_{i,j} \equiv   \delta_i \delta_j .
\end{equation}

Assuming that the $\delta$-field is Gaussian the Fisher matrix for the parameters $\tilde{\pmb{\phi}}$ evaluated at $\tilde{\pmb{\phi}} = 0$ is
\begin{equation}\label{eq:fisher_matrix}
F^{\alpha\beta} = -\left\langle \frac{\partial^2\ln\mathcal{L} }{\partial\tilde{\pmb{\phi}}_\alpha \partial\tilde{\pmb{\phi}}_\beta } \right\rangle = \frac{1}{2} \tr\left[ {\bf P}^\alpha {\bf C}^{-1}  {\bf P^*}^{\beta} {\bf C}^{-1} \right].
\end{equation}
This is the standard result for a multivariant Gaussian and easily derived.

Now consider the quantity
\begin{align}
 \Ave{ \pmb{\delta}^\intercal {\bf C}^{-1} {\bf P^*}^\mu {\bf C}^{-1} \pmb{\delta} } 
& =  \Ave{ \tr\left[ {\bf C}^{-1} {\bf P^*}^\mu {\bf C}^{-1} \pmb{\Delta} \right] } \\
& =   \tr\left[ {\bf C}^{-1} {\bf P^*}^\mu {\bf C}^{-1} \left(  {\bf C}  + {\bf P}^\nu \tilde{\phi}_\nu \right) \right]  \\
&= \tr\left[ {\bf C}^{-1} {\bf P^*}^\mu \right] + \tr\left[ {\bf C}^{-1} {\bf P^*}^\mu {\bf C}^{-1} {\bf P}^\nu \right] \tilde{\phi}_\nu \\
&= \tr\left[ {\bf C}^{-1} {\bf P^*}^\mu \right] + 2 F^{\mu\nu} \tilde{\phi}_\nu
\end{align}
where ~(\ref{eq:fisher_matrix}) and the invariance of the trace under cyclic permutations was used.
We can therefore construct the unbiased estimator that can be written in several different forms
\begin{align} \label{eq:est_final}
\hat{\phi}_\mu  & = \frac{1}{2} F^{-1}_{\mu \nu} \left( \pmb{\delta}^\intercal {\bf C}^{-1} {\bf P^*}^\nu {\bf C}^{-1} \pmb{\delta} 
- \tr\left[ {\bf C}^{-1} {\bf P^*}^\nu \right]  \right)  \\
& = \frac{1}{2}  F^{-1}_{\mu \nu} \left( \tr\left[   {\bf C}^{-1} {\bf P^*}^\nu {\bf C}^{-1} \pmb{\Delta} \right]
- \tr\left[ {\bf C}^{-1} {\bf P^*}^\nu \right]  \right)  \\
& = \frac{1}{2}  F^{-1}_{\mu \nu}  \tr\left[   {\bf C}^{-1} {\bf P^*}^\nu \left(  {\bf C}^{-1} \pmb{\Delta} 
-  \bf{I}  \right)  \right]  \\
& =  \frac{1}{2} F^{-1}_{\mu \nu}
\left( \pmb{z}^\intercal {\bf P^*}^\nu \pmb{z} 
- \tr\left[ {\bf C}^{-1} {\bf P^*}^\nu \right]  \right)
\label{eq:est_finalz}
\end{align}
where $\pmb{z} = {\bf C}^{-1} \pmb{\delta}$

The covariance of this estimator is 
\begin{align} \label{eq:cov_raw}
\Ave{ \hat{\phi}_\alpha \hat{\phi}_\beta^*} 
&= \frac{1}{4} F^{-1}_{\alpha \mu} F^{-1}_{\beta \nu} \left( 
C^{-1}_{ij}  P_{jk}^\mu  C_{km}^{-1} 
C^{-1}_{np}\left(  P_{pq}^\nu \right)^* C_{ql}^{-1} \Ave{\Delta_{mi}\Delta_{ln}} 
\right. \\  & \left.
- \tr\left[ {\bf C}^{-1} {\bf P}^\mu \right]  \tr\left[ {\bf C}^{-1}  {\bf P^*}^\nu  \right] \right)
\end{align}

Now we will assume that the distribution of the pixel values pre-lensing is Gaussian.  This allows us to calculate the fourth order moment
\begin{equation}\label{eq:wick}
 \Ave{\Delta_{mi}\Delta_{ln}} = \left\langle \delta_m \delta_i \delta_l \delta_n \right\rangle = C_{mi} C_{ln} + C_{ml} C_{in} + C_{mn} C_{il}
\end{equation}
Plugging this into  (\ref{eq:cov_raw}) gives
\begin{align} \label{eq:fisheer_error}
\Ave{ \hat{\phi}_\alpha \hat{\phi}_\beta^*}  =  F^{-1}_{\alpha \beta}
\end{align}
This is the Cram\'er-Rao lower limit on the covariance of an unbiased estimator so estimator (\ref{eq:est_final}) is an efficient estimator, i.e. the best that can be done given the assumptions.

In the above we have made the approximation that there is just one $\phi(\btheta)$ (or $\balpha(\btheta)$) field for all pixels.  The deflection field should be a function of the pixel's redshift because of the distance factors and because some lensing of the high redshift pixels will be caused by structures that are at the redshift of the lower redshift pixels  (we  call this {\it self-lensing}).  The above formalism can be easily expanded so that there are a series of potential fields, $\phi(\btheta,z_s)$, 
so that the range of angular size distances for the pixels is explicitly taken into account.  When the data can support more free parameters a slightly more complicated treatment can be done.  For now we will consider the $\phi(\btheta)$ field to be averaged over the redshifts of the \lya\ forest.

To calculate the estimator (\ref{eq:est_final}) one needs to find the inverse of the covariance matrix ${\bf C}$.  The data vector  $\pmb{z} = {\bf C}^{-1} \pmb{\delta}$ could be calculated by solving  ${\bf C} \pmb{z} = \pmb{\delta}$ without finding $ {\bf C}^{-1}$, but it is still needed to calculate the Fisher matrix and bias term.  This can be very numerically burdensome in realistic cases.

\subsection{Unconstrained Modes \& Degeneracies}
\label{sec:degeneracies}

Only differences in the deflection between points appear in the estimator so the potential can only be constrained up to an additive constant and a constant gradient.  Physically the gradient is a uniform shift in the sky which does not change the relative positions of the sources.  

In addition to these degeneracies the data may not constrain some other properties of our reconstruction depending on the expansion and the distribution of sources on the sky.  For example, if the parameters of the expansion each have a limited range of influence on the deflection field, as in a gridded representation of the potential, then regions that have no data in them will be unconstrained, i.e. holes in the map or regions off the edge of the survey.  In the extreme case where the number of model parameters exceeds the number of pairs so that the problem is under determined there will clearly be combinations of parameters that are unconstrained, but it is not necessary for the number of parameters to be very high to have some unconstrained combinations. These unconstrained combinations correspond to eigenvectors of $\mat{F}$ that have zero eigenvalues so they can be  identified. 

The physical degeneracies (the offset and gradient) do not necessarily correspond perfectly to some combination of the expansion parameters.  For example the gradient over a small field cannot be expressed as a finite sum of discrete Fourier modes or spherical harmonics.  In these expansions the above procedure will not remove the gradient.  In these cases large artificial gradients can appear if they are not removed in a separate step.  In other expansions, such as the Legendre expansion used later, the three degenerate modes are easily identified and removed. 

Treating the deflection as a potential field necessarily removes the curl or B-mode component of the deflection field.  This component is expected to be quite small in the weak lensing regime and so we will ignore it here.  If the B-modes ever become of interest, they could be reconstructed by solving for the scalar potential and the pseudo-scalar B-mode potential \citep{1996astro.ph..9149S} in a very similar way to what is done here.

\subsection{Choice of basis}

Since the individual correlations between spectra give limited information on the deflection field it makes sense to try to reconstruct a smoothed version of the deflection field.    Using a grid with a resolution higher than the density of sources would clearly make the problem under determined.  As shown above, one possible basis is that of discrete Fourier modes.   The appearance of an artificial gradient as discussed in section~\ref{sec:degeneracies} is a significant drawback to this basis.

It is also possible to use the spherical harmonic coefficients as free parameters (appendix~\ref{app:sphereical}).    This is probably not advantageous unless the data covers almost all of the sky.  Otherwise there will be many unconstrained combinations of  spherical harmonic modes, making it difficult to obtain a solution for the potential field.  This approach might be advantageous for recovering the power spectrum of the potential, but this will not be considered in this paper.

In what follows we use the Legendre expansion in a square patch on the sky (appendix~\ref{app:legendre_expansion}).  This has the advantages that the degenerate modes can be easily removed,  the deflection is continuous, there is no implicit periodicity and the expansion can be cut off at a particular angular scale as with a DFT.   The Legendre polynomial of order $n$ has $n$ zeros so the effective angular scale for a mode is the field size divided by $n$.

\subsection{Sparse case \& approximation}
\label{app:sparce_case}

If the sources are clumped into groups that have relatively little cross-correlation in the absorption 
and noise between them, the computational cost of the quadratic estimator can be significantly reduced.  In this case the covariance matrix is block-diagonal
\begin{align}
\pmb{C} = \left( \begin{array}{cccc}
\pmb{C}_{(1)} & 0 & \cdots & 0\\
0 & \pmb{C}_{(2)} & \cdots & 0 \\
0 & 0 & \ddots & 0 \\
0 & 0 & \cdots & \pmb{C}_{(n)} 
\end{array} \right)
\end{align}
where $\pmb{C}_{(k)}$ is the covariance matrix for the sources in the $k$th of $n$ groups.   Likewise 
\begin{align}
\pmb{P}^\alpha =  \left( \begin{array}{c}
\pmb{P}^\alpha _{(1)} \\
\pmb{P}^\alpha _{(2)}  \\
\vdots \\
\pmb{P}^\alpha _{(n)} 
\end{array} \right)
\end{align}
The Fisher matrix~(\ref{eq:fisher_matrix}) becomes 
\begin{equation}
F^{\alpha\beta} = \frac{1}{2} \sum_{k=1}^n \tr\left[ {\bf P}_{(k)}^\alpha {\bf C}_{(k)}^{-1} \left( {\bf P}_{(k)}^{\beta}\right)^* {\bf C}_{(k)}^{-1} \right] = \sum_{k=1}^n F^{\alpha\beta}_{(k)}
\end{equation}
and the estimator (\ref{eq:est_final}) becomes
\begin{align}
\hat{\phi}_\mu  
& = \frac{1}{2}  F^{-1}_{\mu \nu} \sum_{k=1}^n \tr\left[   {\bf C}_{(k)}^{-1} {\bf P}_{(k)}^\nu \left(  {\bf C}_{(k)}^{-1} \pmb{\Delta}_{(k)} 
-  \bf{I}  \right)  \right]  \\
& = \frac{1}{2}  F^{-1}_{\mu \nu} \sum_{k=1}^n \tilde{\phi}^{k}_\nu.  \label{eq:binned_estimate}
\end{align}
Calculating this estimator is faster by a factor of $\sim 1/n^2$, uses less memory by a factor of $\sim 1/n^2$ if each term is calculated in series and has less numerical error than in the general case if the sources are evenly distributed among the groups, but if there is significant cross-correlation between the groups the estimator will be less than optimal.  

The groupings could be in positions on the sky and/or in redshift bins.   To reduce computational cost the data could be broken up into redshift bins, the quantities $F^{\alpha\beta}_{(k)}$ and $\tilde{\phi}^{k}_\nu$ calculated, and then combined with equation~(\ref{eq:binned_estimate})  to get an overall estimate of the potential.  This ignores correlations between bins, both intrinsic and in the noise, but might not entail a significant reduction in the signal-to-noise.

\section{Simulation method}
\label{sec:simulation_method}

To simulate the observations and test out implementation of the estimator we need to simulate the absorption and the lensing potential.  Here we describe how we carry this out.

\subsection{Simulating the \lya\ forest}
\label{sec:sim_forest}

We have used two methods for simulating the absorption field.   The first is to go directly from the covariance matrix $\pmb{C}$ for the pixels to a simulated realization.  The correlation matrix is calculated using the methods discussed in appendix~\ref{app:lya_correlation}.  The Cholesky decomposition of the covariance matrix is
\begin{align}
\pmb{C} = \pmb{L}\pmb{L}^T
\end{align}
where $\pmb{L}$ is a lower triangular matrix.  This decomposition is done using the linear algebra package Eigen\footnote{\url{http://eigen.tuxfamily.org}}.  

We generate $n$ normally distributed numbers:
\begin{align}
\delta_i = \pmb{L} x_i
\end{align}
where the $x_i$'s are uncorrelated normally distributed numbers with variance 1, white noise.  
This data set, $\delta_i$, will have the desired covariance and be statistically equivalent to a sample taken from a Gaussian random field .

Another method is to directly simulate a Gaussian random field in 3D and then sample the field at the locations of the pixels.  This would require a much larger array of simulated pixels than data voxels since it must resolve the pixels and contain enough modes to fully represent the fluctuations on the scale of the voxels.  This is very computationally expensive and will give the same results to the extent that the absorption field is Gaussian random.  

\subsection{Simulating the lensing potential}
\label{sec:sim_lensing}

We simulate the lensing potential by generating a two dimensional Gaussian random field that is several times as large as the intended field size using the standard method in Fourier space.  The field is then cropped to the desired size.  This avoids imposing periodic boundary conditions and includes Fourier modes that are large than the field size.  The resolution of the cropped map is 512 x 512.  The deflection is found by finite difference method in configuration space.
The power spectrum for the lensing potential is calculated with the CAMB \citep{Lewis:1999bs}\footnote{\url{http://camb.info}} and CosmoSIS \citep{2015A&C....12...45Z} \footnote{\url{https://bitbucket.org/joezuntz/cosmosis}} software.
As discussed already, the constant and linear modes are not measurable so these are subtracted from the simulated fields before comparing them with the reconstructed field.

\begin{figure}
\center
 \includegraphics[width=1.0\columnwidth]{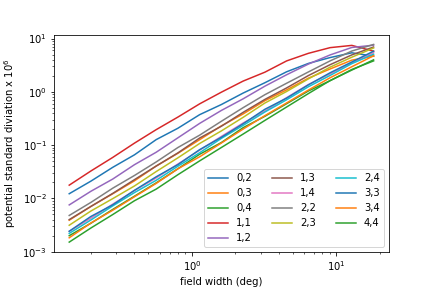}
 \caption{ The standard deviation of the Legendre lensing potential modes (appendix~\ref{app:legendre_expansion}) as a function of field size.  Each standard 
 deviation is estimated using 1000 simulated potential fields. }
 \label{fig:phi_variance}
\end{figure}

The parameters that are actually measured are the Legendre coefficients so the expected signal can be characterized by finding the standard deviation of these coefficients.  This is done by simulating fields and calculating their coefficients by numerical integration.
Figure~\ref{fig:phi_variance} shows the standard deviation of the first 13 Legendre coefficients of 1000 simulations as a function of field size excluding the degenerate modes - (0,0) , (1,0) and (0,1).

\section{Demonstrations}
\label{sec:demonstrations}

\begin{table*}
\caption{Simulation Runs. The number of pixels per spectrum ("\#" in the column headings) is set to 2 for all runs. In addition to the listed parameters the minimum redshift of the \lya\ forest is taken to be $z=2$. $\eta$ is the angular density of sources and $L_{\rm pix}$ is the radial length of the voxels. S/N is the ratio of the expected standard deviation of the coefficients to the standard deviation of the estimator for the coefficients.}
\label{table:simulations}
\centering\begin{tabular}{ccccccccllll}
  \hline
label & number  & field size & $\eta$          & $L_{\rm pix}$  &range of  & number of &                 & max & min \\ 
       & of sources & deg x deg - \# &   deg$^{-2}$ & Mpc  &  modes  & realizations & $\sigma_\delta$ &  S/N & S/N 
       & $\chi^2 / n$ & p-value\\ 
\hline
\hline
\hspace{0.001cm} \\
AA  &  5000  & 1 x 1 - 2 & 5000   & 2  & 5 x 5  & 100 & 0 &  4.2 & 2.3 & 2.8 & $1.3\times 10^{-5}$ \\
DD  &  2000  & 0.5 x 0.5 - 2 & 8000  & 2 & 5 x 5  & 100 & 0 & 3.8 & 2.1 & 5.9 & $\sim 0$ \\
EE &  500  & 0.5 x 0.5 - 2 & 2000     & 2 & 5 x 5  & 100 & 0 & 1.3 &  0.67 & 2.3 & $4.9\times 10^{-4}$ \\
FF &  200  & 0.5 x 0.5 - 2 & 1600     & 2 & 5 x 5  & 100 & 0 & 0.62 &  0.31 & 1.0 & 0.5 \\
CC & 1000 & 1 x 1 - 2 & 1000    &2 & 5 x 5  & 100  & 0 & 1.2 & 0.67 & 2.6 & $5.7\times 10^{-5}$\\
BB  &  5000  & 5 x 5 - 2  & 200    & 2 & 5 x 5  & 100 & 0 & 0.95 & 0.50 & 1.2 & 0.23\\
GG &  2000  & 1 x 1 - 2 & 2000    & 2 & 5 x 5  & 100 & 0.6 &  0.43 &  0.26 & 1.4 & 0.10  \\
HH &  2000  & 1 x 1 - 2 & 2000    & 2 & 5 x 5  & 100 & 0.8 &  0.3 &  0.18  & 0.7 & 0.84\\
II &  2000  & 1 x 1 - 2 & 2000    & 2 & 5 x 5  & 100 & 0.5 &  0.53 & 0.30  & 1.0 & 0.5\\
JJ &  5000  & 1 x 1 - 2 & 5000    & 2 & 5 x 5  & 100 & 0.6 &  1.04 &  0.67 & 2.1 & $1.9\times 10^{-3}$\\
KK &  2000  & 1 x 1 - 2 & 2000   & 1 & 5 x 5  & 200 & 0.6 &  0.62 &  0.36 &  1.0 & 0.5 \\
\hline
\end{tabular}
\end{table*}

\begin{figure*}[!h]%
\subfloat[Simulation AA \label{fig:AA_reconstruction}]{%
\includegraphics[width=\columnwidth,trim=40 80 30 65,clip]{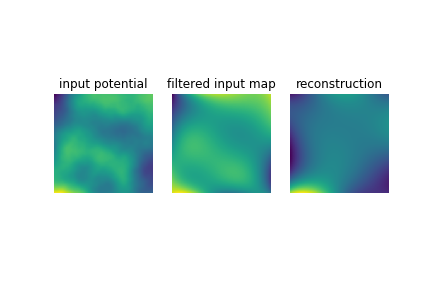}}%
\hfill%
\subfloat[Simulation DD \label{fig:DD_reconstruction}]{%
\includegraphics[width=\columnwidth,trim=40 80 30 65,clip]{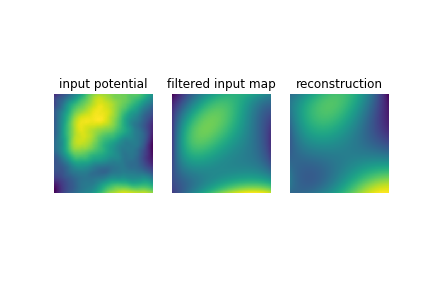}}%
\\%
\subfloat[Simulation EE \label{fig:EE_reconstruction}]{%
\includegraphics[width=\columnwidth,trim=40 80 30 65,clip]{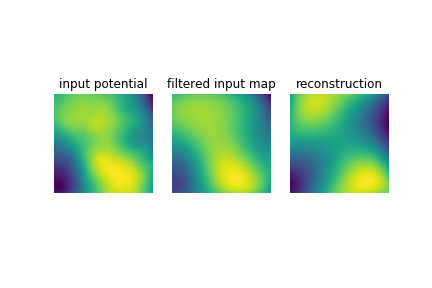}}%
\hfill%
\subfloat[Simulation FF \label{fig:FF_reconstruction}]{%
\includegraphics[width=\columnwidth,trim=40 80 30 65,clip]{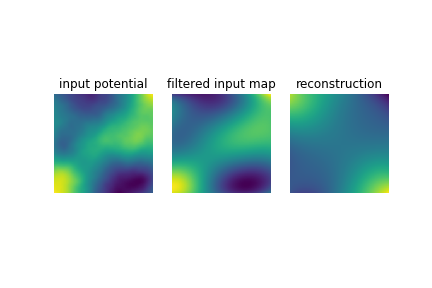}}%
\\%
\subfloat[Simulation CC \label{fig:CC_reconstruction}]{%
\includegraphics[width=\columnwidth,trim=40 80 30 65,clip]{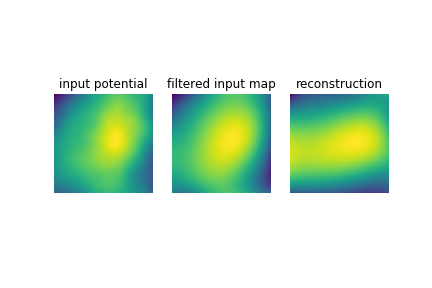}}
\hfill%
\subfloat[Simulation BB \label{fig:BB_reconstruction}]{%
\includegraphics[width=\columnwidth,trim=40 80 30 65,clip]{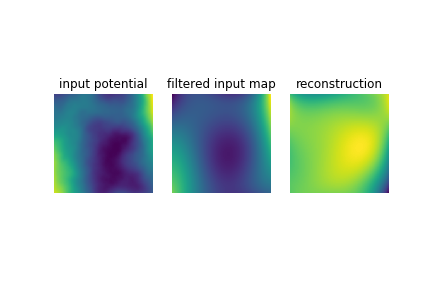}}%
\\%
\subfloat[Simulation GG \label{fig:GG_reconstruction}]{%
\includegraphics[width=\columnwidth,trim=40 80 30 65,clip]{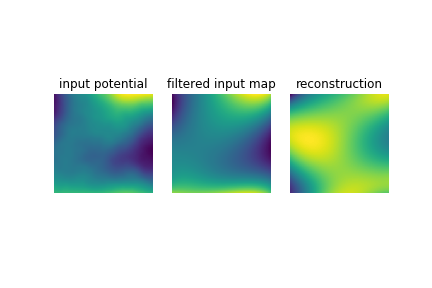}}%
\hfill%
\subfloat[Simulation II \label{fig:II_reconstruction}]{%
\includegraphics[width=\columnwidth,trim=40 80 30 65,clip]{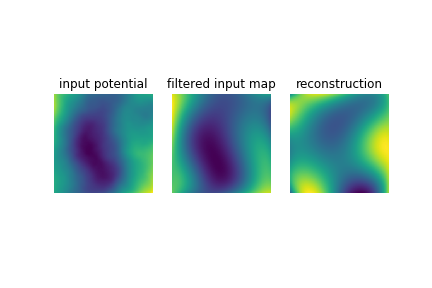}}%
\\%
\subfloat[Simulation JJ \label{fig:JJ_reconstruction}]{%
\includegraphics[width=\columnwidth,trim=40 80 30 65,clip]{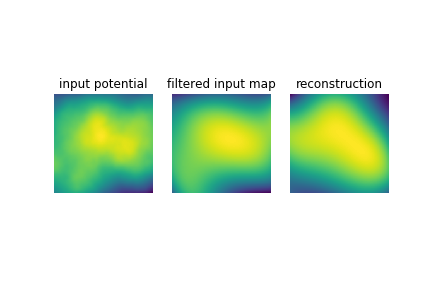}}%
\hfill%
\subfloat[Simulation KK \label{fig:KK_reconstruction}]{%
\includegraphics[width=\columnwidth,trim=40 80 30 65,clip]{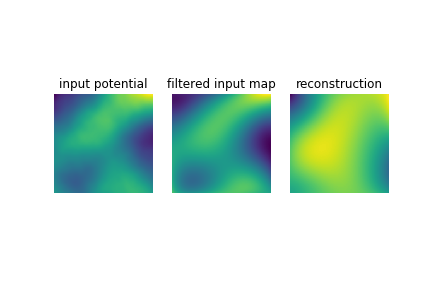}}%
\caption{%
Images of the lensing potential from the simulations of Table~\ref{table:simulations}.
The left panels of each case show the random realization of the input lensing potential with the mean and gradient subtracted.
The centre panels show only the reconstructed modes of the input field.
The right panels show the reconstructed field using 100 (or, for KK, 200) stacked redshift layers.
}%
\clearpage%
\end{figure*}

\begin{figure*}[!h]%
\subfloat[Simulation AA \label{fig:AA_variances}]{%
\includegraphics[width=\columnwidth]{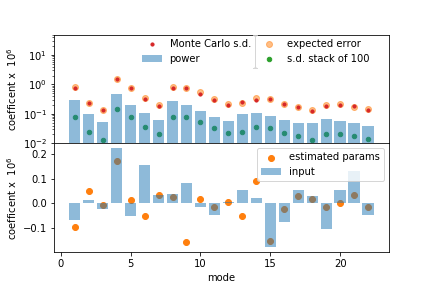}}%
\hfill%
\subfloat[Simulation DD \label{fig:DD_variances}]{%
\includegraphics[width=\columnwidth]{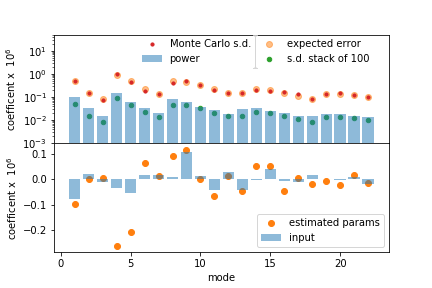}}%
\\%
\subfloat[Simulation EE \label{fig:EE_variances}]{%
\includegraphics[width=\columnwidth]{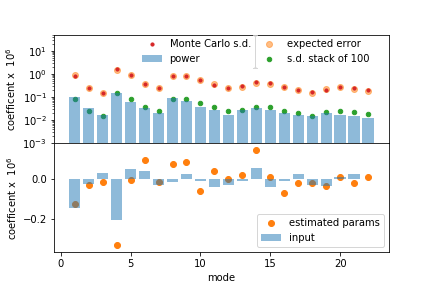}}%
\hfill%
\subfloat[Simulation FF \label{fig:FF_variances}]{%
\includegraphics[width=\columnwidth]{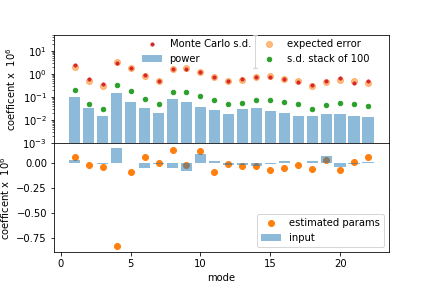}}%
\\%
\subfloat[Simulation CC \label{fig:CC_variances}]{%
\includegraphics[width=\columnwidth]{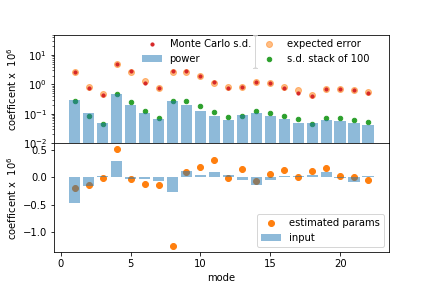}}%
\hfill%
\subfloat[Simulation BB \label{fig:BB_variances}]{%
\includegraphics[width=\columnwidth]{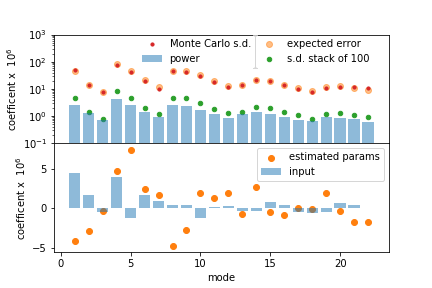}}%
\caption{%
In the top panel of each case are the standard deviations of the Legendre modes of the lensing potential and the expected errors in the estimator.   The blue bars are the variance of the coefficients calculated from a thousand realizations of the lensing potential.   The orange dots in the top panels are the expected standard deviation of the estimate of the coefficients according to the analytic formula  (\ref{eq:fisheer_error}).   The smaller red dots are the variance among the 100 simulations.   They are in good agreement in all cases.  The green dots are the errors expected if 100 (200 for case KK) sets of the simulation are stacked as if they are at different redshifts.  By comparing the green dots to the blue bars some idea of the signal-to-noise can be gained. In this case the noise levels for 200 pixel spectra are below the typical signal values expected.
In the lower panels are the coefficients for the particular realization used in the simulation (blue bars) and the estimated values of those coefficients after stacking 100 (200 for case KK)  realizations of the \lya\ forest. (orange dots).   The modes are listed without the ones that are not measured ($mn = 00, 10$ and $01$).
}%
\clearpage%
\end{figure*}

\begin{figure*}%
\ContinuedFloat%
\subfloat[Simulation GG \label{fig:GG_variances}]{%
\includegraphics[width=\columnwidth]{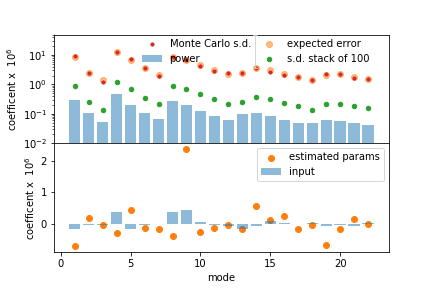}}%
\hfill%
\subfloat[Simulation II \label{fig:II_variances}]{%
\includegraphics[width=\columnwidth]{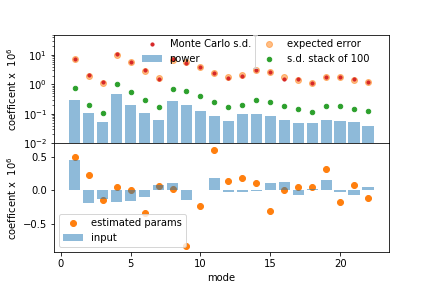}}%
\\%
\subfloat[Simulation JJ \label{fig:JJ_variances}]{%
\includegraphics[width=\columnwidth]{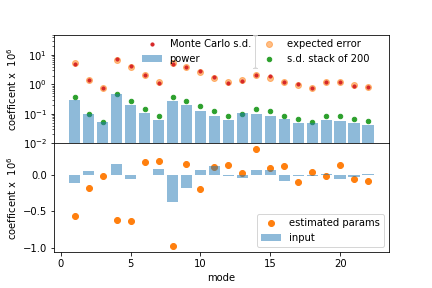}}%
\hfill%
\subfloat[Simulation KK \label{fig:KK_variances}]{%
\includegraphics[width=\columnwidth]{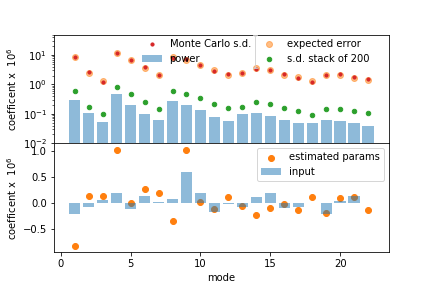}}%
\caption{\emph{Continued}}%
\end{figure*}

\begin{figure}
\center
 \includegraphics[width=1.0\columnwidth]{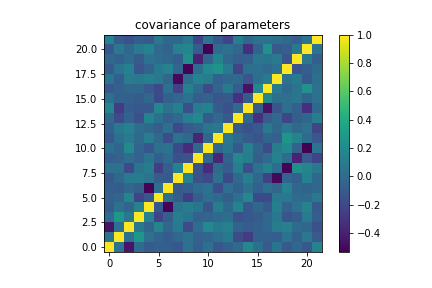}
 \caption{The normalized covariance between the coefficient estimates in simulation run AA (table~\ref{table:simulations}).  The estimates are consistent with being uncorrelated.}
 \label{fig:AA_covariances}
\end{figure}

To test how well the algorithm can reconstruct the lensing potential we carried out a series of simulations.   
The source image positions are chosen randomly within the field.  The lensing potential is simulated as described in section~\ref{sec:sim_lensing} and the deflection at each of the source positions is calculated.  The spectroscopic pixels are displaced according to the deflection and a realization of the absorption in each pixel 
is created as described in section~\ref{sec:sim_forest} and appendix~\ref{app:lya_correlation}.  Then the estimator is applied to these pixels to recover the Legendre coefficients of the lensing potential.  An image of the reconstructed potential is created using these coefficients that can be compared to the original.  This is done many times with new realizations of the \lya\ forest to calculate the variance in the recovered coefficients. 

The parameters that characterize each simulation run are:  1) the angular size of the field over which the potential is reconstructed, 2) the pixel length which is the length of the spectroscopic bin that is used to measure $\delta$ (this could be the spectrograph's pixel size or an average over multiple pixels, but we will still call it a "pixel" for simplicity.)  This length is expressed in comoving Mpc using the the conversion appropriate to the background cosmology.) 3) the number of sources within the field 4) number of pixels in each spectrum (in real data this will vary among sources, but for the simulations it is the same for all sources.) 5) the redshift of the lowest redshift pixel in each spectrum, which is taken to be the same for all sources,  6) the noise in each pixel, $\sigma_\delta$ , which for these tests we take to be uncorrelated between pixels, 7) the number of coefficients in the expansion of the potential that are used.

Spectra used for studying the \lya\ forest typically consist of hundreds of pixels and the number of sources we expect to be in the hundreds to tens of thousands.  This results in $\sim 10^4$ to $\sim 10^{8}$ pixel pairs.  This makes the storage and manipulation of the matrices involved difficult.  Fortunately, as we will show in section~\ref{sec:noise_scaling}, the correlation between pixels at different redshifts (given the adopted pixel lengths) are small and so, for the purposes of these simulations, they can be considered independent.  Using this property, we reconstruct the lensing potential using only two consecutive pixels in each spectrum.  These pixels are at the same redshifts for all the sources.  The \lya \, simulation within this redshift slice contains all the correct correlations between pixels.
The optimal way to combine statistically independent redshift slices is to simply average them (see section~\ref{app:sparce_case}).   Using this property, we find the final potential reconstruction by averaging the reconstructed lens maps from 100 redshift slices, each with different realizations of the forest.  The result should have the same noise properties as the case with 200 pixels in each of the spectra.   We do not simulate gaps in the spectra caused by masking, but this poses no fundamental problem.  This procedure greatly reduces the computational cost.

Table~\ref{table:simulations} gives the parameter values for each of the simulations that will be discussed and some measures of the fidelity of the reconstruction are given in the last four columns.  The signal-to-noise, $S/N$, is calculated by dividing the expected variance in each coefficient due to signal (from Monte Carlo as in figure~\ref{fig:phi_variance}) and by the variance expected for the estimator.  This is done for each mode and the range is given in columns 9 and 10.

The $\chi^2 /n$ column in table~\ref{table:simulations} is calculated for all of the coefficients (22 of them) that are measured for the particular random realization simulated with the null hypothesis being that the data is only the expected noise without any lensing signal.  The last column of of table is the p-value for the  $\chi^2$ in the previous column.  This p-value represents the probability of getting a $\chi^2$ larger than what is measured if there were no lensing signal.  Small values signify stronger detections.

Figure~\ref{fig:AA_reconstruction} shows the input and reconstructed potential fields for case AA, 5000 sources in one square degree.  The initial potential field has structure in it on a smaller scale than is represented by the finite Legendre series that is used in the reconstruction.  In the central panel of Figure~\ref{fig:AA_reconstruction} (and the others like it) we show the potential constructed by calculating only the Legendre coefficients of the input field that we are trying to estimate and filtering the others out.  It can be seen that it lacks some of the detail of the original input fields.  The right hand panel of Figure~\ref{fig:AA_reconstruction} (and the others like it) is the reconstructed field using the estimated Legendre coefficients.    By chance this field has relatively mild structure in the potential.  (No attempt has been made to select better looking cases.)  Never the less, in the AA case the reconstructed image 
broadly reproduces many of the features seen in the input field.  

Figure~\ref{fig:AA_variances}  shows the results for case AA more quantitatively.  The upper panel shows predictions for the average signal and noise in each Legendre coefficient.  It can be seen there that the analytic predictions given in section~\ref{sec:QE} for the variance of the estimator agree well with the variances estimated from the simulations.  It can also be seen that the noise in this case is well below the expected signal.  In the lower panel of figure~\ref{fig:AA_variances}  are the actual input and output values for the particular potential field used in the simulation.  The p-value in table~\ref{table:simulations} is very small indicating the a strong detection would be expected in this case.

Simulation run DD is shown in figure~\ref{fig:DD_reconstruction} and \ref{fig:DD_variances}.  This case is for a smaller area (0.5 x 0.5 deg) with a higher source density as compared to run AA.  The potential seems better recovered than in the AA run, but this appears to be only because the structure in the potential happens to be more robust.  The upper panel of \ref{fig:DD_variances} shows that the signal-to-noise would typically be smaller in this case than for AA although still above one.  Note that the scale in figure~\ref{fig:DD_variances} is different than that in \ref{fig:AA_variances}.  The expected fluctuations on a smaller field are smaller.  The $\chi^2 /n = 5.9$ in this case which gives a p-value that is too small to calculate accurately meaning that the detection would be very strong.

Simulation run EE is for the same size field as DD, but with fewer sources, 500 compared to 2000.  Figure~\ref{fig:EE_variances} shows that the signal-to-noise is smaller, but still apparently good enough to recover the major features in the field shown in figure~\ref{fig:EE_reconstruction} and 
yields a very small p-value.  Run FF is the same, but with even fewer sources, 200.  Here it seems that the potential cannot be accurately recovered (figures~\ref{fig:FF_reconstruction} and \ref{fig:FF_variances}).  The $\chi^2 /n = 1.0$, consistent with noise. 

Run CC is the same as AA, but with one fifth the number of sources.  From figure~\ref{fig:CC_reconstruction} it appears that a local peak in the potential is being detected, but not much more detail is present.  Figure~\ref{fig:CC_variances} confirms the impression that the signal-to-noise is around one or a bit smaller for the individual coefficients.  The p-value very small indicating that a signal is strongly detected.

Run BB is with a larger field than AA and the same number of sources.  Figures~\ref{fig:BB_reconstruction} and \ref{fig:BB_variances}, and the $\chi^2$ show that we should not expect to be able to reconstruct the potential with source densities this low despite the variance of the Legendre coefficients being larger on this scale than it is for smaller fields.  

In the first six simulation runs listed in table~\ref{table:simulations} the pixel noise was taken to be zero.  This represents the optimistic case were the intrinsic variation in the absorption in each pixel is larger than then the noise, signal dominated.  This is not the case for the majority of sources in current surveys that reach the kind of source densities we are considering.   For example CLAMATO \citep{clamato18} and LATIS \citep{2020arXiv200210676N} both have median pixel noise $\sigma_\delta \sim 0.58$ which makes them noise dominated for individual pixels ($\sigma_{\rm IGM}\sim 0.28$ in our case) although the bright quasar sources are signal dominated.   For comparison LATIS expects to have a source density of 2,000 deg$^{-2}$ over 1.7 deg$^2$ and CLAMATO about half the density over 1 deg$^{2}$.  

In the final simulations we added pixel noise that is taken to be constant and uncorrelated.  In this case $\pmb{N} = \sigma_\delta^2 \, \pmb{I}$.  Run GG has 2000 deg$^{-2}$ and $\sigma_\delta = 0.6$.  The results are shown in figures~\ref{fig:GG_reconstruction} and \ref{fig:GG_variances}.  The signal takes a significant hit and although  some structure does seem to be recovered the p-value is 0.1 which signifies a marginal detection.  The noise was reduced somewhat in run II to $\sigma_\delta = 0.5$ (figures~\ref{fig:II_reconstruction} and \ref{fig:II_variances}).  The image of the potential is better recovered, but the signal-to-noise for the individual coefficients is always below one.  The $\chi^2$ is larger than for GG probably due to chance.

Run JJ (figures~\ref{fig:JJ_reconstruction} and \ref{fig:JJ_variances}) has $\sigma_\delta = 0.6$, but 2.5 times higher source density than GG.  The fidelity is much improved and the p-value quite low.  Finally, run KK (figures~\ref{fig:KK_reconstruction} and \ref{fig:KK_variances}) has smaller spectral pixels ($L_{\rm pix} = 1$ Mpc), but the same total range in wavelength as the other runs.  The other parameters are as in case GG.  The fidelity is similarly to GG indicating that the signal-to-noise is not sensitive to $L_{\rm pix}$, at least in this regime.

Figure~\ref{fig:AA_covariances} shows the normalized covariance between estimated Legendre coefficients in case AA.  These are estimated from the repeated simulations with random \lya\ forests.  It can be seen here that the estimates of the different coefficients are not strongly correlated and thus can be considered independent measurements.  This plot looks very similar for all the other cases.

\subsection{Noise scaling}
\label{sec:noise_scaling}


\begin{figure}
\center
\includegraphics[width=1.0\columnwidth]{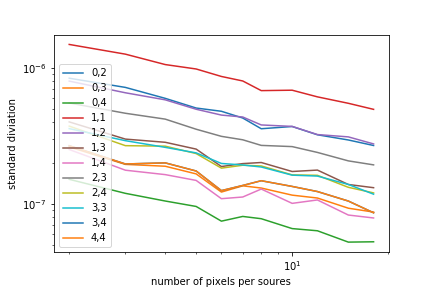}
\caption{
The standard deviation of the Legendre lensing potential modes (appendix~\ref{app:legendre_expansion}) computed using 
 the estimator as a function of the number of pixels
 in each of the spectra for a 0.5 deg x 0.5 deg field with 500 sources.}
\label{fig:error_vs_Nspec}
\end{figure}


\begin{figure}
\center
 \includegraphics[width=1.0\columnwidth]{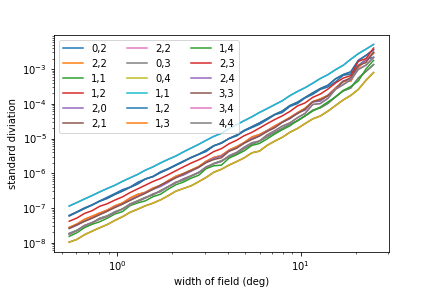}
 \caption{  
 The standard deviation of the Legendre lensing potential modes (appendix~\ref{app:legendre_expansion}) computed using 
 the estimator as a function
  of the width of a square field  with 1000 sources and a depth of 200 pixels.}
 \label{fig:error_vs_length}
\end{figure}

\begin{figure}
\center
 \includegraphics[width=1.0\columnwidth]{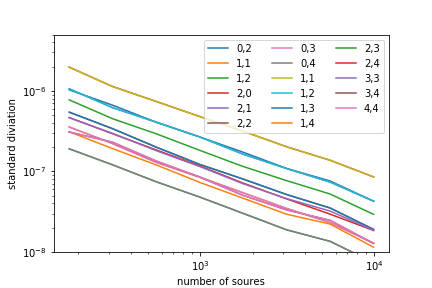}
 \caption{
 The standard deviation of the Legendre lensing potential modes (appendix~\ref{app:legendre_expansion}) computed using 
 the estimator as a function
 number of sources 
 in a  1 deg x 1 deg field.  This is for 200 pixels per source.}
 \label{fig:error_vs_Nqso}
\end{figure}

Figures \ref{fig:error_vs_Nspec},  \ref{fig:error_vs_length} and \ref{fig:error_vs_Nqso} show the standard deviation of the estimator as a function of number of spectral pixels per source, size of the field on the sky and  number of sources respectively.  These are calculated directly from the expected error given in Section~\ref{sec:QE}.  It can be seen in Figure~\ref{fig:error_vs_Nspec} that the variance scales as $N_{\rm spec}^{-1}$ indicating that each map of the \lya\ forest at a constant redshift constitutes essentially an independent measurement of the lensing potential, at least for a pixel length of 2~Mpc.  This means that a calculation of the noise with one or two pixels per source can be easily scaled to find the noise for more pixels and that the potential estimates can be "stacked" as was done in the previous section (Note that Section~\ref{app:sparce_case} shows that the optimal weighting is not the simple average of the estimates except in the cases where the noise,  pixels positions and intrinsic correlations are identical in each redshift bin.)  

Figure~\ref{fig:error_vs_length} shows that the scaling with the size of the field is not quite a power-law.  Nevertheless the error in each coefficient scales approximately as
\begin{align} \label{eq:scaling_relation}
\sigma_{nm} \simpropto N_{\rm spec}^{\gamma}  L^\beta N_{\rm source}^\alpha
\end{align}
where $\gamma \simeq -0.5$, $\beta \simeq 2.6$ and $\alpha \sim 0.8$ for the modes and ranges of parameters that we have investigated.


\section{Sensitivity to assumption about the correlation function}
\label{sec:systematics}

It should be noted that the quadratic estimator contains the correlation function of the \lya\ forest which must be estimated in some way and thus might be in error.  If the assumed covariance $\pmb{C}$ is incorrect the estimator will be biased.  In the simplest case where the assumed correlation function is off by a normalization,  $\pmb{C} = a \Ave{\pmb{\Delta}}$ where $a$ is a constant, the average of the estimator will be
\begin{align}
\Ave{\hat{\phi}_\mu} = a ~\phi_\mu + \frac{(a-1)}{2} ~ F^{-1}_{\mu\nu} \tr\left[ {\bf C}^{-1}  {\bf P^*}^\nu  \right] 
\end{align}
so that there will be an erroneous scaling and bias.  Given that the correlation function can be assumed to be isotropic and a smooth function of distance it seems reasonable that it can be constrained well enough to allow for a good measurement of the lensing.
However this is a problem that has not yet been investigated thoroughly.

\section{Discussion}
\label{sec:discussion}

We have derived a minimum variance estimator for the gravitational lensing potential using the \lya\ forest and found that it can recover an image of the the potential on $\sim 1$ degree scales if the source density is high and the pixel noise is low.  We carried out simulations that showed that our implementation functions as expected.  At the source densities and noise levels currently available
 \citep{clamato18,2020arXiv200210676N} 
the signal-to-noise in the recovered map is expected to be marginal, but with improvements in either of these a high fidelity map could be recovered.

The ELT (Extremely Large Telescope) should be able to reach a magnitude fainter than CLAMATO or LATIS which will mean source densities that are an order of magnitude greater ($\eta \sim 10^4$ deg$^{-3}$) and a reduction in the pixel noise to point where it is subdominant for a much larger fraction of the pixels \citep{2019BAAS...51g..45E,2019BAAS...51g.229S}.  This should be a better case than even our most optimistic simulations so high fidelity lensing maps should be possible.

The estimator proposed here is computationally expensive.  The storage is $\propto N_{pix}^2$ and the computational time $\propto N_{pix}^3$.  This could pose a limitation, but we believe this can be significantly reduced through some methods in development.  A remaining question is whether the correlation function of the \lya\ forest, that is needed for the estimator, is or can be determined to high enough accuracy to not bias the lensing potential estimate.  We do not believe that this posses a significant problem in the long run, but it has still to be determined precisely how well this can be done.

No attempt has been made here to recover the lensing power spectrum from the \lya\ forest.  For the small field surveys studied here any such estimate would be dominated by sample variance, but it might be 
possible for larger, sparser surveys \citep[see][]{2017arXiv170608939M}.  Estimating the power spectrum will require a different approach and will be the subject of a future paper.

The method used by the Planck collaboration to map the lensing using the CMB is essentially to Wiener filter the temperature and polarization maps to fill in masked regions of the sky and then use an estimator based in spherical harmonic space \citep{2014A&A...571A..17P,2016A&A...594A..15P}.  This is not an optimal method and is not applicable to the \lya\ forest because any map of the sky that took full advantage of the accuracy of the source positions would have such high resolution and so few pixels containing sources that it would not be practical.  Some effort has been made to develop a real-space lensing estimator for the CMB \citep{2012PhRvD..85d3016B}.  Our method could be applied to the CMB by treating each unmasked pixel as a source and it would take into account window functions and inhomogeneous noise in an optimal way.   For the whole Planck map this might be computationally onerous, but it is certainly possible for smaller CMB surveys in its present form.



\section*{Acknowledgments}
 RACC was supported by  NASA NNX17AK56G, NASA ATP 80NSSC18K101, NSF AST-1615940, NSF AST-1614853 and  NSF AST-1909193.
\nocite{*}

\bibliographystyle{aa}
\bibliography{references}

\begin{thebibliography}{91}
\expandafter\ifx\csname natexlab\endcsname\relax\def\natexlab#1{#1}\fi

\bibitem[{{Ahn} {et~al.}(2012){Ahn}, {Alexandroff}, {Allende Prieto},
  {Anderson}, {Anderton}, {Andrews}, {Aubourg}, {Bailey}, {Balbinot}, {Barnes},
  {Bautista}, {Beers}, {Beifiori}, {Berlind}, {Bhardwaj}, {Bizyaev}, {Blake},
  {Blanton}, {Blomqvist}, {Bochanski}, {Bolton}, {Borde}, {Bovy}, {Brandt},
  {Brinkmann}, {Brown}, {Brownstein}, {Bundy}, {Busca}, {Carithers}, {Carnero},
  {Carr}, {Casetti-Dinescu}, {Chen}, {Chiappini}, {Comparat}, {Connolly},
  {Crepp}, {Cristiani}, {Croft}, {Cuesta}, {da Costa}, {Davenport}, {Dawson},
  {de Putter}, {De Lee}, {Delubac}, {Dhital}, {Ealet}, {Ebelke}, {Edmondson},
  {Eisenstein}, {Escoffier}, {Esposito}, {Evans}, {Fan}, {Femen{\'\i}a
  Castell{\'a}}, {Fern{\'a}ndez Alvar}, {Ferreira}, {Filiz Ak}, {Finley},
  {Fleming}, {Font-Ribera}, {Frinchaboy}, {Garc{\'\i}a-Hern{\'a}ndez},
  {Garc{\'\i}a P{\'e}rez}, {Ge}, {G{\'e}nova-Santos}, {Gillespie}, {Girardi},
  {Gonz{\'a}lez Hern{\'a}ndez}, {Grebel}, {Gunn}, {Guo}, {Haggard}, {Hamilton},
  {Harris}, {Hawley}, {Hearty}, {Ho}, {Hogg}, {Holtzman}, {Honscheid},
  {Huehnerhoff}, {Ivans}, {Ivezi{\'c}}, {Jacobson}, {Jiang}, {Johansson},
  {Johnson}, {Kauffmann}, {Kirkby}, {Kirkpatrick}, {Klaene}, {Knapp}, {Kneib},
  {Le Goff}, {Leauthaud}, {Lee}, {Lee}, {Long}, {Loomis}, {Lucatello},
  {Lundgren}, {Lupton}, {Ma}, {Ma}, {MacDonald}, {Mack}, {Mahadevan}, {Maia},
  {Majewski}, {Makler}, {Malanushenko}, {Malanushenko}, {Manchado},
  {Mandelbaum}, {Manera}, {Maraston}, {Margala}, {Martell}, {McBride},
  {McGreer}, {McMahon}, {M{\'e}nard}, {Meszaros}, {Miralda-Escud{\'e}},
  {Montero-Dorta}, {Montesano}, {Morrison}, {Muna}, {Munn}, {Murayama},
  {Myers}, {Neto}, {Nguyen}, {Nichol}, {Nidever}, {Noterdaeme}, {Nuza}, {Ogand
  o}, {Olmstead}, {Oravetz}, {Owen}, {Padmanabhan}, {Palanque-Delabrouille},
  {Pan}, {Parejko}, {Parihar}, {P{\^a}ris}, {Pattarakijwanich}, {Pepper},
  {Percival}, {P{\'e}rez-Fournon}, {P{\'e}rez-R{\`a}fols}, {Petitjean},
  {Pforr}, {Pieri}, {Pinsonneault}, {Porto de Mello}, {Prada}, {Price-Whelan},
  {Raddick}, {Rebolo}, {Rich}, {Richards}, {Robin}, {Rocha-Pinto}, {Rockosi},
  {Roe}, {Ross}, {Ross}, {Rossi}, {Rubi{\~n}o-Martin}, {Samushia}, {Sanchez
  Almeida}, {S{\'a}nchez}, {Santiago}, {Sayres}, {Schlegel}, {Schlesinger},
  {Schmidt}, {Schneider}, {Schultheis}, {Schwope}, {Sc{\'o}ccola}, {Seljak},
  {Sheldon}, {Shen}, {Shu}, {Simmerer}, {Simmons}, {Skibba}, {Skrutskie},
  {Slosar}, {Sobreira}, {Sobeck}, {Stassun}, {Steele}, {Steinmetz}, {Strauss},
  {Streblyanska}, {Suzuki}, {Swanson}, {Tal}, {Thakar}, {Thomas}, {Thompson},
  {Tinker}, {Tojeiro}, {Tremonti}, {Vargas Maga{\~n}a}, {Verde}, {Viel},
  {Vikas}, {Vogt}, {Wake}, {Wang}, {Weaver}, {Weinberg}, {Weiner}, {West},
  {White}, {Wilson}, {Wisniewski}, {Wood-Vasey}, {Yanny}, {Y{\`e}che}, {York},
  {Zamora}, {Zasowski}, {Zehavi}, {Zhao}, {Zheng}, {Zhu}, \& {Zinn}}]{ahn12}
{Ahn}, C.~P., {Alexandroff}, R., {Allende Prieto}, C., {et~al.} 2012, \apjs,
  203, 21

\bibitem[{{Ahumada} {et~al.}(2019){Ahumada}, {Allende Prieto}, {Almeida},
  {Anders}, {Anderson}, {Andrews}, {Anguiano}, {Arcodia}, {Armengaud},
  {Aubert}, {Avila}, {Avila-Reese}, {Badenes}, {Balland}, {Barger},
  {Barrera-Ballesteros}, {Basu}, {Bautista}, {Beaton}, {Beers}, {Benavides},
  {Bender}, {Bernardi}, {Bershady}, {Beutler}, {Moni Bidin}, {Bird}, {Bizyaev},
  {Blanc}, {Blanton}, {Boquien}, {Borissova}, {Bovy}, {Brandt}, {Brinkmann},
  {Brownstein}, {Bundy}, {Bureau}, {Burgasser}, {Burtin}, {Cano-Diaz},
  {Capasso}, {Cappellari}, {Carrera}, {Chabanier}, {Chaplin}, {Chapman},
  {Cherinka}, {Chiappini}, {Choi}, {Chojnowski}, {Chung}, {Clerc}, {Coffey},
  {Comerford}, {Comparat}, {da Costa}, {Cousinou}, {Covey}, {Crane}, {Cunha},
  {da Silva Ilha}, {Dai}, {Damsted}, {Darling}, {Horta Darrington}, {Davidson},
  {Davies}, {Dawson}, {De}, {de la Macorra}, {De Lee}, {Queiroz}, {Deconto
  Machado}, {de la Torre}, {Dell'Agli}, {du Mas des Bourboux},
  {Diamond-Stanic}, {Dillon}, {Donor}, {Drory}, {Duckworth}, {Dwelly},
  {Ebelke}, {Eftekharzadeh}, {Davis Eigenbrot}, {Elsworth}, {Eracleous},
  {Erfanianfar}, {Escoffier}, {Fan}, {Farr}, {Fernandez-Trincado}, {Feuillet},
  {Finoguenov}, {Fofie}, {Fraser-McKelvie}, {Frinchaboy}, {Fromenteau}, {Fu},
  {Galbany}, {Garcia}, {Garcia-Hernandez}, {Garma Oehmichen}, {Ge}, {Geimba
  Maia}, {Geisler}, {Gelfand }, {Goddy}, {Le Goff}, {Gonzalez-Perez},
  {Grabowski}, {Green}, {Grier}, {Guo}, {Guy}, {Harding}, {Hasselquist},
  {Hawken}, {Hayes}, {Hearty}, {Hekker}, {Hogg}, {Holtzman}, {Hou}, {Hsieh},
  {Huber}, {Hunt}, {Ider Chitham}, {Imig}, {Jaber}, {Jimenez Angel}, {Johnson},
  {Jones}, {Jonsson}, {Jullo}, {Kim}, {Kinemuchi}, {Kirkpatrick}, {Kite},
  {Klaene}, {Kneib}, {Kollmeier}, {Kong}, {Kounkel}, {Krishnarao}, {Lacerna},
  {Lan}, {Lane}, {Law}, {Leung}, {Lewis}, {Li}, {Lian}, {Lin}, {Long},
  {Longa-Pena}, {Lundgren}, {Lyke}, {Mackereth}, {MacLeod}, {Majewski},
  {Manchado}, {Maraston}, {Martini}, {Masseron}, {Masters}, {Mathur},
  {McDermid}, {Merloni}, {Merrifield}, {Meszaros}, {Miglio}, {Minniti},
  {Minsley}, {Miyaji}, {Gohar Mohammad}, {Mosser}, {Mueller}, {Muna},
  {Munoz-Gutierrez}, {Myers}, {Nadathur}, {Nair}, {Correa do Nascimento},
  {Nevin}, {Newman}, {Nidever}, {Nitschelm}, {Noterdaeme}, {O'Connell},
  {Olmstead}, {Oravetz}, {Oravetz}, {Osorio}, {Pace}, {Padilla},
  {Palanque-Delabrouille}, {Palicio}, {Pan}, {Pan}, {Parker}, {Paviot},
  {Peirani}, {Pena Ramrez}, {Penny}, {Percival}, {Perez-Fournon},
  {Perez-Rafols}, {Petitjean}, {Pieri}, {Pinsonneault}, {Poovelil}, {Povick},
  {Prakash}, {Price-Whelan}, {Raddick}, {Raichoor}, {Ray}, {Barboza Rembold},
  {Rezaie}, {Riffel}, {Riffel}, {Rix}, {Robin}, {Roman-Lopes}, {Roman-Zuniga},
  {Rose}, {Ross}, {Rossi}, {Rowlands}, {Rubin}, {Salvato}, {Sanchez},
  {Sanchez-Menguiano}, {Sanchez-Gallego}, {Sayres}, {Schaefer}, {Schiavon},
  {Schimoia}, {Schlafly}, {Schlegel}, {Schneider}, {Schultheis}, {Schwope},
  {Seo}, {Serenelli}, {Shafieloo}, {Shamsi}, {Shao}, {Shen}, {Shetrone},
  {Shirley}, {Silva Aguirre}, {Simon}, {Skrutskie}, {Slosar}, {Smethurst},
  {Sobeck}, {Cervantes Sodi}, {Souto}, {Stark}, {Stassun}, {Steinmetz},
  {Stello}, {Stermer}, {Storchi-Bergmann}, {Streblyanska}, {Stringfellow},
  {Stutz}, {Suarez}, {Sun}, {Taghizadeh-Popp}, {Talbot}, {Tayar}, {Thakar},
  {Theriault}, {Thomas}, {Thomas}, {Tinker}, {Tojeiro}, {Hernandez Toledo},
  {Tremonti}, {Troup}, {Tuttle}, {Unda-Sanzana}, {Valentini},
  {Vargas-Gonzalez}, {Vargas-Magana}, {Vazquez-Mata}, {Vivek}, {Wake}, {Wang},
  {Weaver}, {Weijmans}, {Wild}, {Wilson}, {Wilson}, {Wolthuis}, {Wood-Vasey},
  {Yan}, {Yang}, {Yeche}, {Zamora}, {Zarrouk}, {Zasowski}, {Zhang}, {Zhao},
  {Zhao}, {Zheng}, {Zheng}, {Zhu}, \& {Zou}}]{ebossdr16}
{Ahumada}, R., {Allende Prieto}, C., {Almeida}, A., {et~al.} 2019, arXiv
  e-prints, arXiv:1912.02905

\bibitem[{{Anderson} {et~al.}(2018){Anderson}, {Luciw}, {Li}, {Kuo}, {Yadav},
  {Masui}, {Chang}, {Chen}, {Oppermann}, {Liao}, {Pen}, {Price},
  {Staveley-Smith}, {Switzer}, {Timbie}, \& {Wolz}}]{anderson18}
{Anderson}, C.~J., {Luciw}, N.~J., {Li}, Y.~C., {et~al.} 2018, \mnras, 476,
  3382

\bibitem[{{Bartelmann} \& {Maturi}(2017)}]{bart17}
{Bartelmann}, M. \& {Maturi}, M. 2017, Scholarpedia, 12, 32440

\bibitem[{{Bartelmann} \& {Schneider}(2001)}]{bart01}
{Bartelmann}, M. \& {Schneider}, P. 2001, \physrep, 340, 291

\bibitem[{{Bautista} {et~al.}(2015){Bautista}, {Bailey}, {Font-Ribera},
  {Pieri}, {Busca}, {Miralda-Escud{\'e}}, {Palanque-Delabrouille}, {Rich},
  {Dawson}, {Feng}, {Ge}, {Gontcho}, {Ho}, {Le Goff}, {Noterdaeme},
  {P{\^a}ris}, {Rossi}, \& {Schlegel}}]{bautista15}
{Bautista}, J.~E., {Bailey}, S., {Font-Ribera}, A., {et~al.} 2015, \jcap, 2015,
  060

\bibitem[{{Bernardeau}(1997)}]{bernard97}
{Bernardeau}, F. 1997, \aap, 324, 15

\bibitem[{{Birrer} {et~al.}(2018){Birrer}, {Refregier}, \& {Amara}}]{birrer18}
{Birrer}, S., {Refregier}, A., \& {Amara}, A. 2018, \apjl, 852, L14

\bibitem[{{Blomqvist} {et~al.}(2019){Blomqvist}, {du Mas des Bourboux},
  {Busca}, {de Sainte Agathe}, {Rich}, {Balland}, {Bautista}, {Dawson},
  {Font-Ribera}, {Guy}, {Le Goff}, {Palanque-Delabrouille}, {Percival},
  {P{\'e}rez-R{\`a}fols}, {Pieri}, {Schneider}, {Slosar}, \&
  {Y{\`e}che}}]{blom19}
{Blomqvist}, M., {du Mas des Bourboux}, H., {Busca}, N.~G., {et~al.} 2019,
  \aap, 629, A86

\bibitem[{{Blomqvist} {et~al.}(2015){Blomqvist}, {Kirkby}, {Bautista},
  {Arinyo-i-Prats}, {Busca}, {Miralda-Escud{\'e}}, {Slosar}, {Font-Ribera},
  {Margala}, {Schneider}, \& {Vazquez}}]{blomqvist15}
{Blomqvist}, M., {Kirkby}, D., {Bautista}, J.~E., {et~al.} 2015, \jcap, 2015,
  034

\bibitem[{{Bolton} {et~al.}(2017){Bolton}, {Puchwein}, {Sijacki}, {Haehnelt},
  {Kim}, {Meiksin}, {Regan}, \& {Viel}}]{bolton17}
{Bolton}, J.~S., {Puchwein}, E., {Sijacki}, D., {et~al.} 2017, \mnras, 464, 897

\bibitem[{{Bordoloi} {et~al.}(2010){Bordoloi}, {Lilly}, \&
  {Amara}}]{bordoloi10}
{Bordoloi}, R., {Lilly}, S.~J., \& {Amara}, A. 2010, \mnras, 406, 881

\bibitem[{{Bucher} {et~al.}(2012){Bucher}, {Carvalho}, {Moodley}, \&
  {Remazeilles}}]{2012PhRvD..85d3016B}
{Bucher}, M., {Carvalho}, C.~S., {Moodley}, K., \& {Remazeilles}, M. 2012,
  \prd, 85, 043016

\bibitem[{{Busca} {et~al.}(2013{\natexlab{a}}){Busca}, {Delubac}, {Rich},
  {Bailey}, {Font-Ribera}, {Kirkby}, {Le Goff}, {Pieri}, {Slosar}, {Aubourg},
  {Bautista}, {Bizyaev}, {Blomqvist}, {Bolton}, {Bovy}, {Brewington}, {Borde},
  {Brinkmann}, {Carithers}, {Croft}, {Dawson}, {Ebelke}, {Eisenstein},
  {Hamilton}, {Ho}, {Hogg}, {Honscheid}, {Lee}, {Lundgren}, {Malanushenko},
  {Malanushenko}, {Margala}, {Maraston}, {Mehta}, {Miralda-Escud{\'e}},
  {Myers}, {Nichol}, {Noterdaeme}, {Olmstead}, {Oravetz},
  {Palanque-Delabrouille}, {Pan}, {P{\^a}ris}, {Percival}, {Petitjean}, {Roe},
  {Rollinde}, {Ross}, {Rossi}, {Schlegel}, {Schneider}, {Shelden}, {Sheldon},
  {Simmons}, {Snedden}, {Tinker}, {Viel}, {Weaver}, {Weinberg}, {White},
  {Y{\`e}che}, \& {York}}]{busca13}
{Busca}, N.~G., {Delubac}, T., {Rich}, J., {et~al.} 2013{\natexlab{a}}, \aap,
  552, A96

\bibitem[{{Busca} {et~al.}(2013{\natexlab{b}})}]{2013A&A...552A..96B}
{Busca}, N.~G. {et~al.} 2013{\natexlab{b}}, \aap, 552, A96

\bibitem[{{Chabanier} {et~al.}(2020){Chabanier}, {Bournaud}, {Dubois},
  {Palanque-Delabrouille}, {Y{\`e}che}, {Armengaud}, \& {Peirani}}]{chab20}
{Chabanier}, S., {Bournaud}, F., {Dubois}, Y., {et~al.} 2020, arXiv e-prints,
  arXiv:2002.02822

\bibitem[{{Chabanier} {et~al.}(2019){Chabanier}, {Millea}, \&
  {Palanque-Delabrouille}}]{chabanier19}
{Chabanier}, S., {Millea}, M., \& {Palanque-Delabrouille}, N. 2019, \mnras,
  489, 2247

\bibitem[{{Chang} {et~al.}(2010){Chang}, {Pen}, {Bandura}, \&
  {Peterson}}]{chang10}
{Chang}, T.-C., {Pen}, U.-L., {Bandura}, K., \& {Peterson}, J.~B. 2010, \nat,
  466, 463

\bibitem[{{Chang} {et~al.}(2004){Chang}, {Refregier}, \& {Helfand}}]{chang04}
{Chang}, T.-C., {Refregier}, A., \& {Helfand}, D.~J. 2004, \apj, 617, 794

\bibitem[{{Cregg} \& {Svedlindh}(2007)}]{2007JPhA...4014029C}
{Cregg}, P.~J. \& {Svedlindh}, P. 2007, Journal of Physics A Mathematical
  General, 40, 14029

\bibitem[{{Croft} {et~al.}(2018){Croft}, {Romeo}, \& {Metcalf}}]{croft18}
{Croft}, R. A.~C., {Romeo}, A., \& {Metcalf}, R.~B. 2018, \mnras, 477, 1814

\bibitem[{{Croft} {et~al.}(1998){Croft}, {Weinberg}, {Katz}, \&
  {Hernquist}}]{croft98}
{Croft}, R. A.~C., {Weinberg}, D.~H., {Katz}, N., \& {Hernquist}, L. 1998,
  \apj, 495, 44

\bibitem[{{Cunha} {et~al.}(2014){Cunha}, {Huterer}, {Lin}, {Busha}, \&
  {Wechsler}}]{cunha14}
{Cunha}, C.~E., {Huterer}, D., {Lin}, H., {Busha}, M.~T., \& {Wechsler}, R.~H.
  2014, \mnras, 444, 129

\bibitem[{{Dalton} {et~al.}(2018){Dalton}, {Trager}, {Abrams}, {Bonifacio},
  {Aguerri}, {Vallenari}, {Middleton}, {Benn}, {Dee}, {Say{\`e}de}, {Lewis},
  {Pragt}, {Pic{\'o}}, {Walton}, {Rey}, {Allende}, {Lhom{\'e}}, {Terrett},
  {Brock}, {Gilbert}, {Ridings}, {Verheijen}, {Tosh}, {Steele}, {Stuik},
  {Kroes}, {Tromp}, {Kragt}, {Lesman}, {Mottram}, {Bates}, {Gribbin}, {Burgal},
  {Herreros}, {Delgado}, {Martin}, {Cano}, {Navarro}, {Irwin}, {Lewis},
  {Gonzales Solares}, {O'Mahony}, {Bianco}, {Zurita}, {ter Horst}, {Molinari},
  {Lodi}, {Guerra}, {Baruffolo}, {Carrasco}, {Farkas}, {Schallig}, {Hill},
  {Smith}, {Drew}, {Poggianti}, {Pieri}, {Jin}, {Dominquez Palmero},
  {Fari{\~n}a}, {Martin}, {Worley}, {Murphy}, {Hidalgo}, {Mignot}, {Bishop},
  {Guest}, {Elswijk}, {de Haan}, {Hanenburg}, {Salasnich}, {Mayya},
  {Izazaga-P{\'e}rez}, \& {Peralta de Arriba}}]{dalton18}
{Dalton}, G., {Trager}, S., {Abrams}, D.~C., {et~al.} 2018, in Society of
  Photo-Optical Instrumentation Engineers (SPIE) Conference Series, Vol. 10702,
  \procspie, 107021B

\bibitem[{{de Sainte Agathe} {et~al.}(2019){de Sainte Agathe}, {Balland}, {du
  Mas des Bourboux}, {Busca}, {Blomqvist}, {Guy}, {Rich}, {Font-Ribera},
  {Pieri}, {Bautista}, {Dawson}, {Le Goff}, {de la Macorra},
  {Palanque-Delabrouille}, {Percival}, {P{\'e}rez-R{\`a}fols}, {Schneider},
  {Slosar}, \& {Y{\`e}che}}]{ebosslyabao}
{de Sainte Agathe}, V., {Balland}, C., {du Mas des Bourboux}, H., {et~al.}
  2019, \aap, 629, A85

\bibitem[{{DESI Collaboration} {et~al.}(2016){DESI Collaboration}, {Aghamousa},
  {Aguilar}, {Ahlen}, {Alam}, {Allen}, {Allende Prieto}, {Annis}, {Bailey},
  {Balland}, {Ballester}, {Baltay}, {Beaufore}, {Bebek}, {Beers}, {Bell},
  {Bernal}, {Besuner}, {Beutler}, {Blake}, {Bleuler}, {Blomqvist}, {Blum},
  {Bolton}, {Briceno}, {Brooks}, {Brownstein}, {Buckley-Geer}, {Burden},
  {Burtin}, {Busca}, {Cahn}, {Cai}, {Cardiel-Sas}, {Carlberg}, {Carton},
  {Casas}, {Castand er}, {Cervantes-Cota}, {Claybaugh}, {Close}, {Coker},
  {Cole}, {Comparat}, {Cooper}, {Cousinou}, {Crocce}, {Cuby}, {Cunningham},
  {Davis}, {Dawson}, {de la Macorra}, {De Vicente}, {Delubac}, {Derwent},
  {Dey}, {Dhungana}, {Ding}, {Doel}, {Duan}, {Ealet}, {Edelstein},
  {Eftekharzadeh}, {Eisenstein}, {Elliott}, {Escoffier}, {Evatt}, {Fagrelius},
  {Fan}, {Fanning}, {Farahi}, {Farihi}, {Favole}, {Feng}, {Fernandez},
  {Findlay}, {Finkbeiner}, {Fitzpatrick}, {Flaugher}, {Flender}, {Font-Ribera},
  {Forero-Romero}, {Fosalba}, {Frenk}, {Fumagalli}, {Gaensicke}, {Gallo},
  {Garcia-Bellido}, {Gaztanaga}, {Pietro Gentile Fusillo}, {Gerard},
  {Gershkovich}, {Giannantonio}, {Gillet}, {Gonzalez-de-Rivera},
  {Gonzalez-Perez}, {Gott}, {Graur}, {Gutierrez}, {Guy}, {Habib}, {Heetderks},
  {Heetderks}, {Heitmann}, {Hellwing}, {Herrera}, {Ho}, {Holland}, {Honscheid},
  {Huff}, {Hutchinson}, {Huterer}, {Hwang}, {Illa Laguna}, {Ishikawa},
  {Jacobs}, {Jeffrey}, {Jelinsky}, {Jennings}, {Jiang}, {Jimenez}, {Johnson},
  {Joyce}, {Jullo}, {Juneau}, {Kama}, {Karcher}, {Karkar}, {Kehoe}, {Kennamer},
  {Kent}, {Kilbinger}, {Kim}, {Kirkby}, {Kisner}, {Kitanidis}, {Kneib},
  {Koposov}, {Kovacs}, {Koyama}, {Kremin}, {Kron}, {Kronig}, {Kueter-Young},
  {Lacey}, {Lafever}, {Lahav}, {Lambert}, {Lampton}, {Land riau}, {Lang},
  {Lauer}, {Le Goff}, {Le Guillou}, {Le Van Suu}, {Lee}, {Lee}, {Leitner},
  {Lesser}, {Levi}, {L'Huillier}, {Li}, {Liang}, {Lin}, {Linder}, {Loebman},
  {Luki{\'c}}, {Ma}, {MacCrann}, {Magneville}, {Makarem}, {Manera}, {Manser},
  {Marshall}, {Martini}, {Massey}, {Matheson}, {McCauley}, {McDonald},
  {McGreer}, {Meisner}, {Metcalfe}, {Miller}, {Miquel}, {Moustakas}, {Myers},
  {Naik}, {Newman}, {Nichol}, {Nicola}, {Nicolati da Costa}, {Nie}, {Niz},
  {Norberg}, {Nord}, {Norman}, {Nugent}, {O'Brien}, {Oh}, {Olsen}, {Padilla},
  {Padmanabhan}, {Padmanabhan}, {Palanque-Delabrouille}, {Palmese},
  {Pappalardo}, {P{\^a}ris}, {Park}, {Patej}, {Peacock}, {Peiris}, {Peng},
  {Percival}, {Perruchot}, {Pieri}, {Pogge}, {Pollack}, {Poppett}, {Prada},
  {Prakash}, {Probst}, {Rabinowitz}, {Raichoor}, {Ree}, {Refregier}, {Regal},
  {Reid}, {Reil}, {Rezaie}, {Rockosi}, {Roe}, {Ronayette}, {Roodman}, {Ross},
  {Ross}, {Rossi}, {Rozo}, {Ruhlmann-Kleider}, {Rykoff}, {Sabiu}, {Samushia},
  {Sanchez}, {Sanchez}, {Schlegel}, {Schneider}, {Schubnell}, {Secroun},
  {Seljak}, {Seo}, {Serrano}, {Shafieloo}, {Shan}, {Sharples}, {Sholl},
  {Shourt}, {Silber}, {Silva}, {Sirk}, {Slosar}, {Smith}, {Smoot}, {Som},
  {Song}, {Sprayberry}, {Staten}, {Stefanik}, {Tarle}, {Sien Tie}, {Tinker},
  {Tojeiro}, {Valdes}, {Valenzuela}, {Valluri}, {Vargas-Magana}, {Verde},
  {Walker}, {Wang}, {Wang}, {Weaver}, {Weaverdyck}, {Wechsler}, {Weinberg},
  {White}, {Yang}, {Yeche}, {Zhang}, {Zhao}, {Zheng}, {Zhou}, {Zhou}, {Zhu},
  {Zou}, \& {Zu}}]{desi16}
{DESI Collaboration}, {Aghamousa}, A., {Aguilar}, J., {et~al.} 2016, arXiv
  e-prints, arXiv:1611.00036

\bibitem[{{Ellis} \& {Dawson}(2019)}]{2019BAAS...51g..45E}
{Ellis}, R. \& {Dawson}, K. 2019, in BAAS, Vol.~51, 45

\bibitem[{{Faucher-Gigu{\`e}re} {et~al.}(2008){Faucher-Gigu{\`e}re},
  {Prochaska}, {Lidz}, {Hernquist}, \& {Zaldarriaga}}]{faucher08}
{Faucher-Gigu{\`e}re}, C.-A., {Prochaska}, J.~X., {Lidz}, A., {Hernquist}, L.,
  \& {Zaldarriaga}, M. 2008, \apj, 681, 831

\bibitem[{{Flagey} {et~al.}(2018){Flagey}, {McConnachie}, {Szeto}, {Hall},
  {Hill}, \& {Hervieu}}]{mse18}
{Flagey}, N., {McConnachie}, A., {Szeto}, K., {et~al.} 2018, in Society of
  Photo-Optical Instrumentation Engineers (SPIE) Conference Series, Vol. 10704,
  \procspie, 107040V

\bibitem[{{Font-Ribera} {et~al.}(2012){Font-Ribera}, {Miralda-Escud{\'e}},
  {Arnau}, {Carithers}, {Lee}, {Noterdaeme}, {P{\^a}ris}, {Petitjean}, {Rich},
  {Rollinde}, {Ross}, {Schneider}, {White}, \& {York}}]{font2012}
{Font-Ribera}, A., {Miralda-Escud{\'e}}, J., {Arnau}, E., {et~al.} 2012, \jcap,
  2012, 059

\bibitem[{{Foreman} {et~al.}(2018){Foreman}, {Meerburg}, {van Engelen}, \&
  {Meyers}}]{2018JCAP...07..046F}
{Foreman}, S., {Meerburg}, P.~D., {van Engelen}, A., \& {Meyers}, J. 2018,
  \jcap, 7, 046

\bibitem[{{Furlanetto} {et~al.}(2006){Furlanetto}, {Oh}, \& {Briggs}}]{furl06}
{Furlanetto}, S.~R., {Oh}, S.~P., \& {Briggs}, F. 2006, Phys. Rep., 433, 181

\bibitem[{Gradshteyn \& Ryzhik(2007)}]{gradshteyn2007}
Gradshteyn, I.~S. \& Ryzhik, I.~M. 2007, Table of integrals, series, and
  products, seventh edn. (Elsevier/Academic Press, Amsterdam), xlviii+1171,
  translated from the Russian, Translation edited and with a preface by Alan
  Jeffrey and Daniel Zwillinger, With one CD-ROM (Windows, Macintosh and UNIX)

\bibitem[{{Hamilton}(1993)}]{1993ApJ...417...19H}
{Hamilton}, A.~J.~S. 1993, \apj, 417, 19

\bibitem[{{Hanson} {et~al.}(2010){Hanson}, {Challinor}, \& {Lewis}}]{hanson10}
{Hanson}, D., {Challinor}, A., \& {Lewis}, A. 2010, General Relativity and
  Gravitation, 42, 2197

\bibitem[{{Harrison} {et~al.}(2016){Harrison}, {Camera}, {Zuntz}, \&
  {Brown}}]{harrison16}
{Harrison}, I., {Camera}, S., {Zuntz}, J., \& {Brown}, M.~L. 2016, \mnras, 463,
  3674

\bibitem[{{Hilbert} {et~al.}(2007{\natexlab{a}}){Hilbert}, {Metcalf}, \&
  {White}}]{HMW07}
{Hilbert}, S., {Metcalf}, R.~B., \& {White}, S.~D.~M. 2007{\natexlab{a}},
  \mnras, 382, 1494

\bibitem[{{Hilbert} {et~al.}(2007{\natexlab{b}}){Hilbert}, {Metcalf}, \&
  {White}}]{hilbert07}
{Hilbert}, S., {Metcalf}, R.~B., \& {White}, S.~D.~M. 2007{\natexlab{b}},
  \mnras, 382, 1494

\bibitem[{{Hu} \& {Okamoto}(2002)}]{huok02}
{Hu}, W. \& {Okamoto}, T. 2002, \apj, 574, 566

\bibitem[{{Hui} {et~al.}(1999){Hui}, {Stebbins}, \& {Burles}}]{hui99}
{Hui}, L., {Stebbins}, A., \& {Burles}, S. 1999, \apjl, 511, L5

\bibitem[{{Ir{\v{s}}i{\v{c}}} {et~al.}(2017){Ir{\v{s}}i{\v{c}}}, {Viel},
  {Haehnelt}, {Bolton}, \& {Becker}}]{irsic17}
{Ir{\v{s}}i{\v{c}}}, V., {Viel}, M., {Haehnelt}, M.~G., {Bolton}, J.~S., \&
  {Becker}, G.~D. 2017, \prl, 119, 031302

\bibitem[{{Jalilvand} {et~al.}(2018){Jalilvand}, {Majerotto}, {Durrer}, \&
  {Kunz}}]{2018arXiv180701351J}
{Jalilvand}, M., {Majerotto}, E., {Durrer}, R., \& {Kunz}, M. 2018, ArXiv
  e-prints [\eprint[arXiv]{1807.01351}]

\bibitem[{{Jeffrey} {et~al.}(2020){Jeffrey}, {Lanusse}, {Lahav}, \&
  {Starck}}]{niall20}
{Jeffrey}, N., {Lanusse}, F., {Lahav}, O., \& {Starck}, J.-L. 2020, \mnras,
  492, 5023

\bibitem[{{Kaiser}(1992)}]{kaiser92}
{Kaiser}, N. 1992, \apj, 388, 272

\bibitem[{{Kaiser} \& {Squires}(1993)}]{kaiser93}
{Kaiser}, N. \& {Squires}, G. 1993, \apj, 404, 441

\bibitem[{{Kovetz} {et~al.}(2017){Kovetz}, {Viero}, {Lidz}, {Newburgh},
  {Rahman}, {Switzer}, {Kamionkowski}, {Aguirre}, {Alvarez}, {Bock}, {Bond},
  {Bower}, {Bradford}, {Breysse}, {Bull}, {Chang}, {Cheng}, {Chung}, {Cleary},
  {Corray}, {Crites}, {Croft}, {Dor{\'e}}, {Eastwood}, {Ferrara}, {Fonseca},
  {Jacobs}, {Keating}, {Lagache}, {Lakhlani}, {Liu}, {Moodley}, {Murray},
  {P{\'e}nin}, {Popping}, {Pullen}, {Reichers}, {Saito}, {Saliwanchik},
  {Santos}, {Somerville}, {Stacey}, {Stein}, {Villaescusa-Navarro}, {Visbal},
  {Weltman}, {Wolz}, \& {Zemcov}}]{kovetz17}
{Kovetz}, E.~D., {Viero}, M.~P., {Lidz}, A., {et~al.} 2017, arXiv e-prints,
  arXiv:1709.09066

\bibitem[{{Lai} {et~al.}(2006){Lai}, {Lidz}, {Hernquist}, \&
  {Zaldarriaga}}]{2006ApJ...644...61L}
{Lai}, K., {Lidz}, A., {Hernquist}, L., \& {Zaldarriaga}, M. 2006, \apj, 644,
  61

\bibitem[{{Lee} {et~al.}(2018){Lee}, {Krolewski}, {White}, {Schlegel},
  {Nugent}, {Hennawi}, {M{\"u}ller}, {Pan}, {Prochaska}, {Font-Ribera},
  {Suzuki}, {Glazebrook}, {Kacprzak}, {Kartaltepe}, {Koekemoer}, {Le
  F{\`e}vre}, {Lemaux}, {Maier}, {Nanayakkara}, {Rich}, {Sanders}, {Salvato},
  {Tasca}, \& {Tran}}]{clamato18}
{Lee}, K.-G., {Krolewski}, A., {White}, M., {et~al.} 2018, \apjs, 237, 31

\bibitem[{{Lewis} \& {Challinor}(2006)}]{lewis06}
{Lewis}, A. \& {Challinor}, A. 2006, \physrep, 429, 1

\bibitem[{Lewis {et~al.}(2000)Lewis, Challinor, \& Lasenby}]{Lewis:1999bs}
Lewis, A., Challinor, A., \& Lasenby, A. 2000, \apj, 538, 473

\bibitem[{{Madau} {et~al.}(1997){Madau}, {Meiksin}, \& {Rees}}]{madau97}
{Madau}, P., {Meiksin}, A., \& {Rees}, M.~J. 1997, \apj, 475, 429

\bibitem[{{McDonald}(2003)}]{2003ApJ...585...34M}
{McDonald}, P. 2003, \apj, 585, 34

\bibitem[{{McDonald} \& {Miralda-Escud{\'e}}(1999)}]{mcdonald99}
{McDonald}, P. \& {Miralda-Escud{\'e}}, J. 1999, \apj, 518, 24

\bibitem[{{McQuinn} \& {White}(2011)}]{2011MNRAS.415.2257M}
{McQuinn}, M. \& {White}, M. 2011, \mnras, 415, 2257

\bibitem[{{McQuinn} \& {White}(2015)}]{2015MNRAS.448.1403M}
{McQuinn}, M. \& {White}, M. 2015, \mnras, 448, 1403

\bibitem[{{Metcalf} {et~al.}(2017){Metcalf}, {Croft}, \&
  {Romeo}}]{2017arXiv170608939M}
{Metcalf}, R.~B., {Croft}, R.~A.~C., \& {Romeo}, A. 2017, ArXiv e-prints
  [\eprint[arXiv]{1706.08939}]

\bibitem[{{Metcalf} {et~al.}(2018){Metcalf}, {Croft}, \& {Romeo}}]{metcalf18}
{Metcalf}, R.~B., {Croft}, R. A.~C., \& {Romeo}, A. 2018, \mnras, 477, 2841

\bibitem[{{Metcalf} \& {Silk}(1997)}]{1997ApJ...489....1M}
{Metcalf}, R.~B. \& {Silk}, J. 1997, \apj, 489, 1

\bibitem[{{Metcalf} \& {Silk}(1998)}]{1998ApJ...492L...1M}
{Metcalf}, R.~B. \& {Silk}, J. 1998, \apjl, 492, L1

\bibitem[{{Metcalf} \& {White}(2007{\natexlab{a}})}]{metcalf07}
{Metcalf}, R.~B. \& {White}, S.~D.~M. 2007{\natexlab{a}}, \mnras, 381, 447

\bibitem[{{Metcalf} \& {White}(2007{\natexlab{b}})}]{metcalf&white2006}
{Metcalf}, R.~B. \& {White}, S.~D.~M. 2007{\natexlab{b}}, \mnras, 381, 447

\bibitem[{{Metcalf} \& {White}(2009)}]{metcalf&white2007}
{Metcalf}, R.~B. \& {White}, S.~D.~M. 2009, \mnras, 394, 704

\bibitem[{{Miralda-Escude}(1991)}]{miralda91}
{Miralda-Escude}, J. 1991, \apj, 380, 1

\bibitem[{Neves {et~al.}(2006)Neves, Padilha, Fontes, Rodriguez, Cruz, Barbosa,
  \& Cesar}]{0305-4470-39-18-L06}
Neves, A. A.~R., Padilha, L.~A., Fontes, A., {et~al.} 2006, Journal of Physics
  A: Mathematical and General, 39, L293

\bibitem[{{Newman} {et~al.}(2020){Newman}, {Rudie}, {Blanc}, {Kelson},
  {Rhoades}, {Hare}, {P{\'e}rez}, {Benson}, {Dressler}, {Gonzalez},
  {Kollmeier}, {Konidaris}, {Mulchaey}, {Rauch}, {Le F{\`e}vre}, {Lemaux},
  {Cucciati}, \& {Lilly}}]{2020arXiv200210676N}
{Newman}, A.~B., {Rudie}, G.~C., {Blanc}, G.~A., {et~al.} 2020, arXiv e-prints,
  arXiv:2002.10676

\bibitem[{{Peeples} {et~al.}(2010){Peeples}, {Weinberg}, {Dav{\'e}}, {Fardal},
  \& {Katz}}]{peeples10}
{Peeples}, M.~S., {Weinberg}, D.~H., {Dav{\'e}}, R., {Fardal}, M.~A., \&
  {Katz}, N. 2010, \mnras, 404, 1281

\bibitem[{{Petkova} {et~al.}(2014){Petkova}, {Metcalf}, \&
  {Giocoli}}]{2014MNRAS.445.1954P}
{Petkova}, M., {Metcalf}, R.~B., \& {Giocoli}, C. 2014, \mnras, 445, 1954

\bibitem[{{Planck Collaboration} {et~al.}(2014){Planck Collaboration}, {Ade},
  {Aghanim}, {Armitage-Caplan}, {Arnaud}, {Ashdown}, {Atrio-Barand ela},
  {Aumont}, {Baccigalupi}, {Banday}, {Barreiro}, {Bartlett}, {Basak},
  {Battaner}, {Benabed}, {Beno{\^\i}t}, {Benoit-L{\'e}vy}, {Bernard},
  {Bersanelli}, {Bielewicz}, {Bobin}, {Bock}, {Bonaldi}, {Bonavera}, {Bond},
  {Borrill}, {Bouchet}, {Bridges}, {Bucher}, {Burigana}, {Butler}, {Cardoso},
  {Catalano}, {Challinor}, {Chamballu}, {Chiang}, {Chiang}, {Christensen},
  {Church}, {Clements}, {Colombi}, {Colombo}, {Couchot}, {Coulais}, {Crill},
  {Curto}, {Cuttaia}, {Danese}, {Davies}, {Davis}, {de Bernardis}, {de Rosa},
  {de Zotti}, {D{\'e}chelette}, {Delabrouille}, {Delouis}, {D{\'e}sert},
  {Dickinson}, {Diego}, {Dole}, {Donzelli}, {Dor{\'e}}, {Douspis}, {Dunkley},
  {Dupac}, {Efstathiou}, {En{\ss}lin}, {Eriksen}, {Finelli}, {Forni},
  {Frailis}, {Franceschi}, {Galeotta}, {Ganga}, {Giard}, {Giardino},
  {Giraud-H{\'e}raud}, {Gonz{\'a}lez-Nuevo}, {G{\'o}rski}, {Gratton},
  {Gregorio}, {Gruppuso}, {Gudmundsson}, {Hansen}, {Hanson}, {Harrison},
  {Henrot-Versill{\'e}}, {Hern{\'a}ndez-Monteagudo}, {Herranz}, {Hildebrand t},
  {Hivon}, {Ho}, {Hobson}, {Holmes}, {Hornstrup}, {Hovest}, {Huffenberger},
  {Jaffe}, {Jaffe}, {Jones}, {Juvela}, {Keih{\"a}nen}, {Keskitalo}, {Kisner},
  {Kneissl}, {Knoche}, {Knox}, {Kunz}, {Kurki-Suonio}, {Lagache},
  {L{\"a}hteenm{\"a}ki}, {Lamarre}, {Lasenby}, {Laureijs}, {Lavabre},
  {Lawrence}, {Leahy}, {Leonardi}, {Le{\'o}n-Tavares}, {Lesgourgues}, {Lewis},
  {Liguori}, {Lilje}, {Linden-V{\o}rnle}, {L{\'o}pez-Caniego}, {Lubin},
  {Mac{\'\i}as-P{\'e}rez}, {Maffei}, {Maino}, {Mand olesi}, {Mangilli},
  {Maris}, {Marshall}, {Martin}, {Mart{\'\i}nez-Gonz{\'a}lez}, {Masi},
  {Massardi}, {Matarrese}, {Matthai}, {Mazzotta}, {Melchiorri}, {Mendes},
  {Mennella}, {Migliaccio}, {Mitra}, {Miville-Desch{\^e}nes}, {Moneti},
  {Montier}, {Morgante}, {Mortlock}, {Moss}, {Munshi}, {Murphy}, {Naselsky},
  {Nati}, {Natoli}, {Netterfield}, {N{\o}rgaard-Nielsen}, {Noviello},
  {Novikov}, {Novikov}, {Osborne}, {Oxborrow}, {Paci}, {Pagano}, {Pajot},
  {Paoletti}, {Partridge}, {Pasian}, {Patanchon}, {Perdereau}, {Perotto},
  {Perrotta}, {Piacentini}, {Piat}, {Pierpaoli}, {Pietrobon}, {Plaszczynski},
  {Pointecouteau}, {Polenta}, {Ponthieu}, {Popa}, {Poutanen}, {Pratt},
  {Pr{\'e}zeau}, {Prunet}, {Puget}, {Pullen}, {Rachen}, {Rebolo}, {Reinecke},
  {Remazeilles}, {Renault}, {Ricciardi}, {Riller}, {Ristorcelli}, {Rocha},
  {Rosset}, {Roudier}, {Rowan-Robinson}, {Rubi{\~n}o-Mart{\'\i}n}, {Rusholme},
  {Sandri}, {Santos}, {Savini}, {Scott}, {Seiffert}, {Shellard}, {Smith},
  {Spencer}, {Starck}, {Stolyarov}, {Stompor}, {Sudiwala}, {Sunyaev}, {Sureau},
  {Sutton}, {Suur-Uski}, {Sygnet}, {Tauber}, {Tavagnacco}, {Terenzi},
  {Toffolatti}, {Tomasi}, {Tristram}, {Tucci}, {Tuovinen}, {Umana},
  {Valenziano}, {Valiviita}, {Van Tent}, {Vielva}, {Villa}, {Vittorio}, {Wade},
  {Wandelt}, {White}, {White}, {Yvon}, {Zacchei}, \&
  {Zonca}}]{2014A&A...571A..17P}
{Planck Collaboration}, {Ade}, P.~A.~R., {Aghanim}, N., {et~al.} 2014, \aap,
  571, A17

\bibitem[{{Planck Collaboration} {et~al.}(2016){Planck Collaboration}, {Ade},
  {Aghanim}, {Arnaud}, {Ashdown}, {Aumont}, {Baccigalupi}, {Banday},
  {Barreiro}, {Bartlett}, {Bartolo}, {Basak}, {Battaner}, {Benabed},
  {Beno{\^\i}t}, {Benoit-L{\'e}vy}, {Bernard}, {Bersanelli}, {Bielewicz},
  {Bock}, {Bonaldi}, {Bonavera}, {Bond}, {Borrill}, {Bouchet}, {Boulanger},
  {Bucher}, {Burigana}, {Butler}, {Calabrese}, {Cardoso}, {Catalano},
  {Challinor}, {Chamballu}, {Chiang}, {Christensen}, {Church}, {Clements},
  {Colombi}, {Colombo}, {Combet}, {Couchot}, {Coulais}, {Crill}, {Curto},
  {Cuttaia}, {Danese}, {Davies}, {Davis}, {de Bernardis}, {de Rosa}, {de
  Zotti}, {Delabrouille}, {D{\'e}sert}, {Diego}, {Dole}, {Donzelli},
  {Dor{\'e}}, {Douspis}, {Ducout}, {Dunkley}, {Dupac}, {Efstathiou}, {Elsner},
  {En{\ss}lin}, {Eriksen}, {Fergusson}, {Finelli}, {Forni}, {Frailis},
  {Fraisse}, {Franceschi}, {Frejsel}, {Galeotta}, {Galli}, {Ganga}, {Giard},
  {Giraud-H{\'e}raud}, {Gjerl{\o}w}, {Gonz{\'a}lez-Nuevo}, {G{\'o}rski},
  {Gratton}, {Gregorio}, {Gruppuso}, {Gudmundsson}, {Hansen}, {Hanson},
  {Harrison}, {Henrot-Versill{\'e}}, {Hern{\'a}ndez-Monteagudo}, {Herranz},
  {Hildebrandt}, {Hivon}, {Hobson}, {Holmes}, {Hornstrup}, {Hovest},
  {Huffenberger}, {Hurier}, {Jaffe}, {Jaffe}, {Jones}, {Juvela},
  {Keih{\"a}nen}, {Keskitalo}, {Kisner}, {Kneissl}, {Knoche}, {Kunz},
  {Kurki-Suonio}, {Lagache}, {L{\"a}hteenm{\"a}ki}, {Lamarre}, {Lasenby},
  {Lattanzi}, {Lawrence}, {Leonardi}, {Lesgourgues}, {Levrier}, {Lewis},
  {Liguori}, {Lilje}, {Linden-V{\o}rnle}, {L{\'o}pez-Caniego}, {Lubin},
  {Mac{\'\i}as-P{\'e}rez}, {Maggio}, {Maino}, {Mand olesi}, {Mangilli},
  {Maris}, {Martin}, {Mart{\'\i}nez-Gonz{\'a}lez}, {Masi}, {Matarrese},
  {McGehee}, {Meinhold}, {Melchiorri}, {Mendes}, {Mennella}, {Migliaccio},
  {Mitra}, {Miville-Desch{\^e}nes}, {Moneti}, {Montier}, {Morgante},
  {Mortlock}, {Moss}, {Munshi}, {Murphy}, {Naselsky}, {Nati}, {Natoli},
  {Netterfield}, {N{\o}rgaard-Nielsen}, {Noviello}, {Novikov}, {Novikov},
  {Oxborrow}, {Paci}, {Pagano}, {Pajot}, {Paoletti}, {Pasian}, {Patanchon},
  {Perdereau}, {Perotto}, {Perrotta}, {Pettorino}, {Piacentini}, {Piat},
  {Pierpaoli}, {Pietrobon}, {Plaszczynski}, {Pointecouteau}, {Polenta}, {Popa},
  {Pratt}, {Pr{\'e}zeau}, {Prunet}, {Puget}, {Rachen}, {Reach}, {Rebolo},
  {Reinecke}, {Remazeilles}, {Renault}, {Renzi}, {Ristorcelli}, {Rocha},
  {Rosset}, {Rossetti}, {Roudier}, {Rowan-Robinson}, {Rubi{\~n}o-Mart{\'\i}n},
  {Rusholme}, {Sandri}, {Santos}, {Savelainen}, {Savini}, {Scott}, {Seiffert},
  {Shellard}, {Spencer}, {Stolyarov}, {Stompor}, {Sudiwala}, {Sunyaev},
  {Sutton}, {Suur-Uski}, {Sygnet}, {Tauber}, {Terenzi}, {Toffolatti}, {Tomasi},
  {Tristram}, {Tucci}, {Tuovinen}, {Valenziano}, {Valiviita}, {Van Tent},
  {Vielva}, {Villa}, {Wade}, {Wandelt}, {Wehus}, {White}, {Yvon}, {Zacchei}, \&
  {Zonca}}]{2016A&A...594A..15P}
{Planck Collaboration}, {Ade}, P.~A.~R., {Aghanim}, N., {et~al.} 2016, \aap,
  594, A15

\bibitem[{{Pourtsidou} \& {Metcalf}(2014{\natexlab{a}})}]{2014MNRAS.439L..36P}
{Pourtsidou}, A. \& {Metcalf}, R.~B. 2014{\natexlab{a}}, \mnras, 439, L36

\bibitem[{{Pourtsidou} \& {Metcalf}(2014{\natexlab{b}})}]{pour14}
{Pourtsidou}, A. \& {Metcalf}, R.~B. 2014{\natexlab{b}}, \mnras, 439, L36

\bibitem[{{Pourtsidou} \& {Metcalf}(2015)}]{2015MNRAS.448.2368P}
{Pourtsidou}, A. \& {Metcalf}, R.~B. 2015, \mnras, 448, 2368

\bibitem[{{Prochaska}(2019)}]{prochaska19}
{Prochaska}, J.~X. 2019, Saas-Fee Advanced Course, 46, 111

\bibitem[{{Rauch}(1998)}]{rauch98}
{Rauch}, M. 1998, \araa, 36, 267

\bibitem[{{Rogers} {et~al.}(2018){Rogers}, {Bird}, {Peiris}, {Pontzen},
  {Font-Ribera}, \& {Leistedt}}]{rogers18}
{Rogers}, K.~K., {Bird}, S., {Peiris}, H.~V., {et~al.} 2018, \mnras, 474, 3032

\bibitem[{{Salvador} {et~al.}(2019){Salvador}, {S{\'a}nchez}, {Pagul},
  {Garc{\'\i}a-Bellido}, {Sanchez}, {Pujol}, {Frieman}, {Gaztanaga}, {Ross},
  {Sevilla-Noarbe}, {Abbott}, {Allam}, {Annis}, {Avila}, {Bertin}, {Brooks},
  {Burke}, {Carnero Rosell}, {Carrasco Kind}, {Carretero}, {Castander},
  {Cunha}, {De Vicente}, {Diehl}, {Doel}, {Evrard}, {Fosalba}, {Gruen},
  {Gruendl}, {Gschwend}, {Gutierrez}, {Hartley}, {Hollowood}, {James}, {Kuehn},
  {Kuropatkin}, {Lahav}, {Lima}, {March}, {Marshall}, {Menanteau}, {Miquel},
  {Romer}, {Roodman}, {Scarpine}, {Schindler}, {Smith}, {Soares-Santos},
  {Sobreira}, {Suchyta}, {Swanson}, {Tarle}, {Thomas}, {Vikram}, {Walker}, \&
  {DES Collaboration}}]{salvador19}
{Salvador}, A.~I., {S{\'a}nchez}, F.~J., {Pagul}, A., {et~al.} 2019, \mnras,
  482, 1435

\bibitem[{{Schaan} \& {Ferraro}(2019)}]{schaan19}
{Schaan}, E. \& {Ferraro}, S. 2019, \prl, 122, 181301

\bibitem[{{Schlegel} {et~al.}(2019){Schlegel}, {Kollmeier}, \&
  {Ferraro}}]{2019BAAS...51g.229S}
{Schlegel}, D., {Kollmeier}, J.~A., \& {Ferraro}, S. 2019, in BAAS, Vol.~51,
  229

\bibitem[{{Schneider} {et~al.}(1992){Schneider}, {Ehlers}, \& {Falco}}]{SEF92}
{Schneider}, P., {Ehlers}, J., \& {Falco}, E.~E. 1992, Gravitational Lenses
  (Springer-Verlag)

\bibitem[{{Slosar} {et~al.}(2013{\natexlab{a}}){Slosar}, {Ir{\v{s}}i{\v{c}}},
  {Kirkby}, {Bailey}, {Busca}, {Delubac}, {Rich}, {Aubourg}, {Bautista},
  {Bhardwaj}, {Blomqvist}, {Bolton}, {Bovy}, {Brownstein}, {Carithers},
  {Croft}, {Dawson}, {Font-Ribera}, {Le Goff}, {Ho}, {Honscheid}, {Lee},
  {Margala}, {McDonald}, {Medolin}, {Miralda-Escud{\'e}}, {Myers}, {Nichol},
  {Noterdaeme}, {Palanque-Delabrouille}, {P{\^a}ris}, {Petitjean}, {Pieri},
  {Pi{\v{s}}kur}, {Roe}, {Ross}, {Rossi}, {Schlegel}, {Schneider}, {Suzuki},
  {Sheldon}, {Seljak}, {Viel}, {Weinberg}, \& {Y{\`e}che}}]{slosar13}
{Slosar}, A., {Ir{\v{s}}i{\v{c}}}, V., {Kirkby}, D., {et~al.}
  2013{\natexlab{a}}, \jcap, 2013, 026

\bibitem[{{Slosar} {et~al.}(2011)}]{2011JCAP...09..001S}
{Slosar}, A. {et~al.} 2011, \jcap, 9, 001

\bibitem[{{Slosar} {et~al.}(2013{\natexlab{b}})}]{2013JCAP...04..026S}
{Slosar}, A. {et~al.} 2013{\natexlab{b}}, \jcap, 4, 026

\bibitem[{{Stebbins}(1996)}]{1996astro.ph..9149S}
{Stebbins}, A. 1996, ArXiv Astrophysics e-prints [\eprint{astro-ph/9609149}]

\bibitem[{{Troxel} {et~al.}(2018){Troxel}, {MacCrann}, {Zuntz}, {Eifler},
  {Krause}, {Dodelson}, {Gruen}, {Blazek}, {Friedrich}, {Samuroff}, {Prat},
  {Secco}, {Davis}, {Fert{\'e}}, {DeRose}, {Alarcon}, {Amara}, {Baxter},
  {Becker}, {Bernstein}, {Bridle}, {Cawthon}, {Chang}, {Choi}, {De Vicente},
  {Drlica-Wagner}, {Elvin-Poole}, {Frieman}, {Gatti}, {Hartley}, {Honscheid},
  {Hoyle}, {Huff}, {Huterer}, {Jain}, {Jarvis}, {Kacprzak}, {Kirk}, {Kokron},
  {Krawiec}, {Lahav}, {Liddle}, {Peacock}, {Rau}, {Refregier}, {Rollins},
  {Rozo}, {Rykoff}, {S{\'a}nchez}, {Sevilla-Noarbe}, {Sheldon}, {Stebbins},
  {Varga}, {Vielzeuf}, {Wang}, {Wechsler}, {Yanny}, {Abbott}, {Abdalla},
  {Allam}, {Annis}, {Bechtol}, {Benoit-L{\'e}vy}, {Bertin}, {Brooks},
  {Buckley-Geer}, {Burke}, {Carnero Rosell}, {Carrasco Kind}, {Carretero},
  {Castander}, {Crocce}, {Cunha}, {D'Andrea}, {da Costa}, {DePoy}, {Desai},
  {Diehl}, {Dietrich}, {Doel}, {Fernandez}, {Flaugher}, {Fosalba},
  {Garc{\'\i}a-Bellido}, {Gaztanaga}, {Gerdes}, {Giannantonio}, {Goldstein},
  {Gruendl}, {Gschwend}, {Gutierrez}, {James}, {Jeltema}, {Johnson}, {Johnson},
  {Kent}, {Kuehn}, {Kuhlmann}, {Kuropatkin}, {Li}, {Lima}, {Lin}, {Maia},
  {March}, {Marshall}, {Martini}, {Melchior}, {Menanteau}, {Miquel}, {Mohr},
  {Neilsen}, {Nichol}, {Nord}, {Petravick}, {Plazas}, {Romer}, {Roodman},
  {Sako}, {Sanchez}, {Scarpine}, {Schindler}, {Schubnell}, {Smith}, {Smith},
  {Soares-Santos}, {Sobreira}, {Suchyta}, {Swanson}, {Tarle}, {Thomas},
  {Tucker}, {Vikram}, {Walker}, {Weller}, {Zhang}, \& {DES
  Collaboration}}]{troxel18}
{Troxel}, M.~A., {MacCrann}, N., {Zuntz}, J., {et~al.} 2018, \prd, 98, 043528

\bibitem[{{Uhlemann} {et~al.}(2018){Uhlemann}, {Pichon}, {Codis}, {L'Huillier},
  {Kim}, {Bernardeau}, {Park}, \& {Prunet}}]{uhle18}
{Uhlemann}, C., {Pichon}, C., {Codis}, S., {et~al.} 2018, \mnras, 477, 2772

\bibitem[{{Valdes} {et~al.}(1983){Valdes}, {Tyson}, \& {Jarvis}}]{valdes83}
{Valdes}, F., {Tyson}, J.~A., \& {Jarvis}, J.~F. 1983, \apj, 271, 431

\bibitem[{{Wyithe} \& {Morales}(2007)}]{wyithe07}
{Wyithe}, J. S.~B. \& {Morales}, M.~F. 2007, \mnras, 379, 1647

\bibitem[{{Zahn} \& {Zaldarriaga}(2006{\natexlab{a}})}]{zz06}
{Zahn}, O. \& {Zaldarriaga}, M. 2006{\natexlab{a}}, \apj, 653, 922

\bibitem[{{Zahn} \& {Zaldarriaga}(2006{\natexlab{b}})}]{ZandZ2006}
{Zahn}, O. \& {Zaldarriaga}, M. 2006{\natexlab{b}}, \apj, 653, 922

\bibitem[{{Zaldarriaga} \& {Seljak}(1999)}]{zald99}
{Zaldarriaga}, M. \& {Seljak}, U. 1999, \prd, 59, 123507

\bibitem[{{Zuntz} {et~al.}(2015){Zuntz}, {Paterno}, {Jennings}, {Rudd},
  {Manzotti}, {Dodelson}, {Bridle}, {Sehrish}, \&
  {Kowalkowski}}]{2015A&C....12...45Z}
{Zuntz}, J., {Paterno}, M., {Jennings}, E., {et~al.} 2015, Astronomy and
  Computing, 12, 45

\end{thebibliography}

\appendix

\section{Expansions}
\label{app:expansions}

In this appendix are the details for several possible expansions or parameterizations of the lensing potential.

\subsection{Legendre expansion}
\label{app:legendre_expansion}

The two dimensional Legendre expansion of the potential will be 
\begin{align}
\phi(\btheta) = \sum_{m=0}^{N_x} \sum_{n=0}^{N_y} a_{mn} F_{nm}(\btheta)
\end{align}
where
\begin{align}
F_{nm}(\btheta) = P_m(x) P_n((y) 
\end{align}
$P_m(x)$ are the Legendre polynomials and
\begin{align}
x \equiv 2\, \frac{\theta_1 - \theta_1^o}{\Delta\theta_x} - 1 ~~~~~~~ y \equiv 2 \,\frac{\theta_2 - \theta_2^o}{\Delta\theta_y} - 1
\end{align}
where $(\theta_1^o,\theta_2^o)$ are the coordinates of the lower left hand corner of the reconstructed field.  $\Delta\theta_x$ and $\Delta\theta_y$ are the width and height of the field.  All the sources are within this field. 

The orthogonality relation for these polynomials results in 
\begin{align}
a_{mn} = \frac{(2n+1)(2m+1)}{4}  \int^1_{-1} dx  \int^1_{-1} dy ~ \phi(x,y)   P_m(x) P_n(y) 
\end{align}
The recursion relation
\begin{align}
\begin{array}{lcl}
P_0(x) &=& 1 \\
P_1(x) &=& x \\
P_{n+1}(x) &=&  \left[  ( 2 n + 1) x P_n(x) - n P_{n-1}(x) \right] / (n+1)
\end{array}
\end{align}
is used to calculate the value of the $P_n(x)$ and the relation
\begin{align}
\frac{d}{dx} P_n(x) = P'_n(x)= \frac{n}{(x^2 -1)} \left[ x P_n(x) - P_{n-1}(x) \right]
\end{align}
is used to calculate its derivative.

Using (\ref{eq:dT}), the deflection is
\begin{align}
\balpha(\btheta) = \nabla \phi(\btheta) =  \sum_{m=0}^{N_x} \sum_{n=0}^{N_y} 2  a_{mn} 
\left( 
\begin{array}{c}
P_n'(x) P_m(y) / \Delta\theta_x \\
P_n(x) P_m'(y)   / \Delta\theta_y
\end{array}
\right)
\end{align}

Some of the advantages of using the Legendre expansion are: 1) No periodic boundary conditions are not implicitly imposed, 2) Unlike bilinear interpolation from a gridded potential, the deflection field is always continuous, 3) Unlike all the previous expansions, the degenerate constant and linear components of the potential field are completely contained in three coefficients 
, $a_{00}$, $a_{01}$ and $a_{10}$, that can be easily removed.

\subsection{Chebyshev expansion}
\label{app:chebyshev_expansion}

The Chebyshev polynomials have the same advantages as the Legendre polynomials as an expansion for the potential.  They generally have more structure near the boundaries for a limited number of modes.

The Chebyshev polynomials of the first kind are given by
the recursion relation
\begin{align}
\begin{array}{lcl}
T_0(x) &=& 1 \\
T_1(x) &=& x \\
T_{n+1}(x) &=& 2 x T_n(x) - T_{n-1}(x)
\end{array}
\end{align}
for the range $-1 \leq x \leq 1$.  The derivative of them is given by
\begin{align}\label{eq:dT}
\frac{d}{dx} T_n(x) = n U_{n-1}(x)
\end{align}
where $U_n(x)$ are the Chebyshev polynomials of the second kind are given by
\begin{align}
\begin{array}{lcl}
U_0(x) &=& 1 \\
U_1(x) &=& 2x \\
U_{n+1}(x) &= &2 x U_n(x) - U_{n-1}(x)
\end{array}
\end{align}

The two dimensional Chebyshev expansion is defined like in the Legendre case
\begin{align}
\phi(\btheta) = \sum_{m=0}^{N_x} \sum_{n=0}^{N_y} a_{mn} F_{nm}(\btheta)
\end{align}
where
\begin{align}
F_{nm}(\btheta) = T_m(x) T_n((y) .
\end{align}
Using (\ref{eq:dT}), the deflection is
\begin{align}
\balpha(\btheta) = \nabla \phi(\btheta) =  \sum_{m=0}^{N_x} \sum_{n=0}^{N_y} 2 a_{mn} 
\left( 
\begin{array}{c}
m U_{m-1} T_n  / \Delta\theta_x \\
n T_m U_{n-1}   / \Delta\theta_y
\end{array}
\right)
\end{align}

Using the orthogonality and normalization of these polynomials we can  find the coefficients  given a potential field
\begin{align}
a_{mn} = \epsilon_m\epsilon_n  \int^1_{-1} dx  \int^1_{-1} dy ~\frac{ \phi(x,y)   T_m(x) T_n(y) }{ \sqrt{1-x^2} \sqrt{1-y^2}}
\end{align}
\begin{align}
\epsilon_i = \frac{1}{\pi} \times \left\{
\begin{array}{cl}
1 & i = 0 \\
2 & i > 0
\end{array}
 \right.
\end{align}

As in the Legendre case, modes $a_{00}$, $a_{01}$ and $a_{10}$ are not measurable because they represent a constant potential and two linear, or constant deflection, components.
\subsection{Discrete Fourier Transform}
\label{app:DFT}
The potential field in the flat-sky approximation can be expanded as a $N_1 \times N_2$ pixel map~$\phi[n_1, n_2]$, $0 \le n_i \le N_i - 1$, using the Discrete Fourier Transform (DFT),
\begin{equation}\label{eq:dft}
    \phi[\vec{n}]
    = \sum_{k_1 = 0}^{N_1 - 1} \sum_{k_2 = 0}^{N_2 - 1} \hat{\phi}_{\vec{k}} \, e^{2\pi i \, k_1 n_1/N_1 + 2\pi i \, k_2 n_2/N_2} \;,
\end{equation}
where $\vec{0} \le \vec{n} < \vec{N}$, $\vec{0} \le \vec{k} < \vec{N}$ are multi-indices, and $\hat{\phi}_{\vec{k}}$ are the $N_1 \times N_2$ complex-valued modes of the expansion.
The discrete field values correspond to the continuous field at the grid points $\vec{\theta}_{\vec{n}} \equiv \vec{n} \Delta$, where $\Delta$ is the pixel size of the discrete map.
We can interpolate the DFT~\eqref{eq:dft} to obtain a continuous potential field over the field of view,
\begin{equation}\label{eq:phi-dft}
    \phi(\vec{\theta})
    = \sum_{k_1 = 0}^{N_1 - 1} \sum_{k_2 = 0}^{N_2 - 1} \hat{\phi}_{\vec{k}} \, e^{i \, \vec{\theta} \cdot \vec{l}_{\vec{k}}} \;,
\end{equation}
where the frequencies $\vec{l}_{\vec{k}} = (l_{k_1}, l_{k_2})$ are chosen to produce a real-valued interpolation,
\begin{equation}\label{eq:dft-freq}
    l_{k_i} \equiv \frac{2\pi}{N_i \Delta} \begin{cases}
        k_i \;,       & 2 k_i < N_i \;, \\
        k_i - N_i \;, & 2 k_i > N_i \;, \\
        0 \;,         & 2 k_i = N_i \;.
    \end{cases}
\end{equation}
The gradient of the potential field expansion~\eqref{eq:phi-dft} is the DFT deflection angle,
\begin{equation}\label{eq:alpha-dft}
    \vec{\alpha}(\vec{\theta})
    = \sum_{k_1 = 0}^{N_1 - 1} \sum_{k_2 = 0}^{N_2 - 1} i \, \vec{l}_{\vec{k}} \, \hat{\phi}_{\vec{k}} \, e^{i \, \vec{\theta} \cdot \vec{l}_{\vec{k}}} \;.
\end{equation}
Inserting this into equation~\eqref{eq:def_P} yields the elements of matrix~$\mat{P}$ for the DFT,
\begin{equation}\label{eq:P-dft}
    P_{ij,\vec{k}}
    \equiv 2 \, \tfrac{\vec{l}_{\vec{k}} \cdot \vec{\Delta\theta}_{ij}}{|\vec{\Delta\theta}_{ij}|} \sin\Bigl(\tfrac{\vec{l}_{\vec{k}} \cdot \vec{\Delta\theta}_{ij}}{2}\Bigr) \, e^{i \, \vec{\vec{l}_{\vec{k}} \cdot \bar{\theta}}_{ij}} \, \mathcal{G}_{ij} \;,
\end{equation}
where $\vec{\Delta\theta}_{ij} \equiv \vec{\theta}_i - \vec{\theta}_j$ and $\vec{\bar{\theta}}_{ij} \equiv (\vec{\theta}_i + \vec{\theta}_j)/2$ are the difference and average of positions $\vec{\theta}_i, \vec{\theta}_j$, respectively.

\begin{figure}[t!]%
\centering%
\includegraphics[width=\columnwidth]{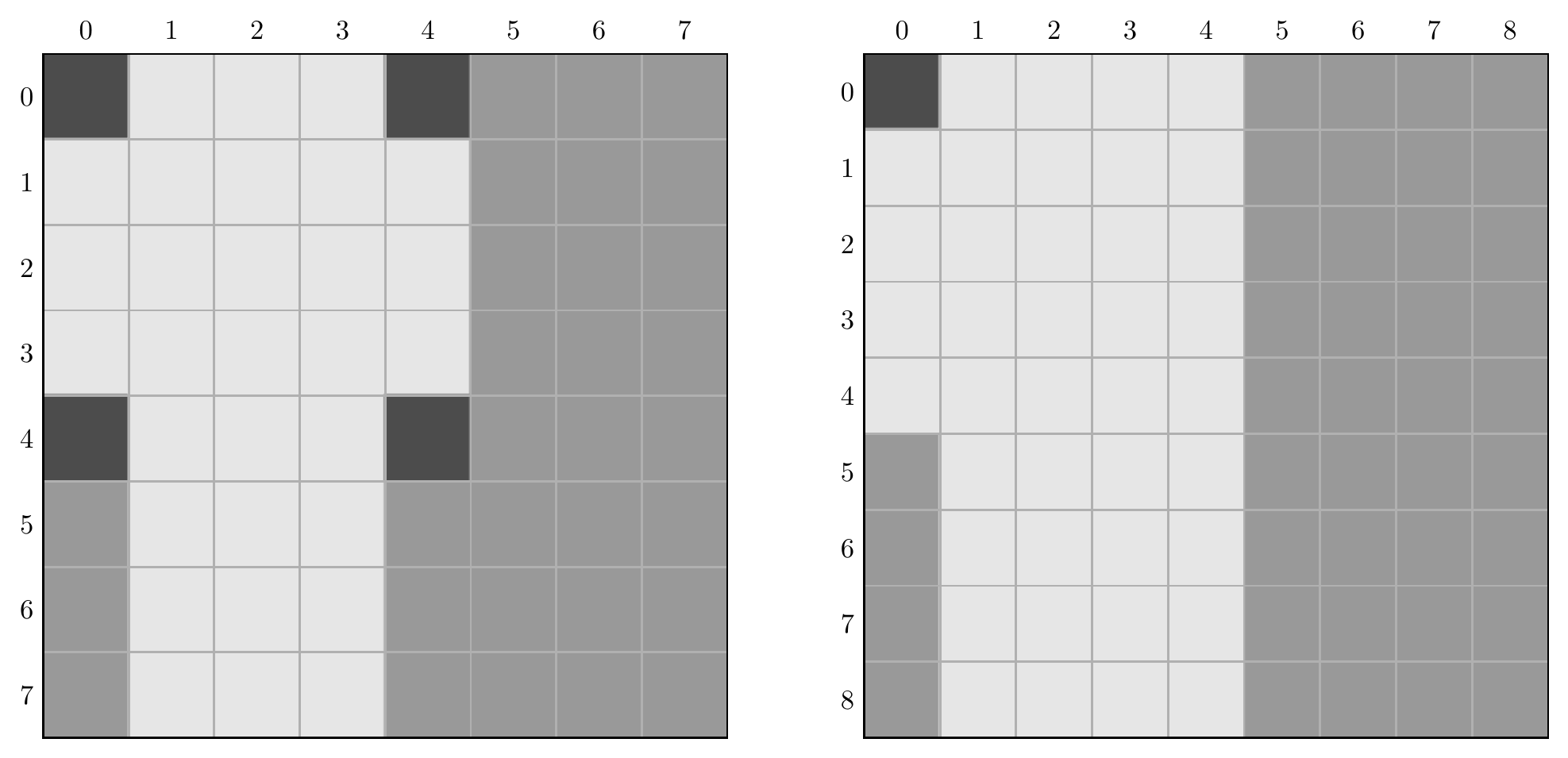}%
\caption{%
The modes in a $8 \times 8$ (\emph{left}) and $9 \times 9$ (\emph{right}) DFT expansion.
Light shaded modes are excluded due to symmetry, dark shaded modes are excluded due to vanishing frequency.
}%
\label{fig:dft}%
\end{figure}

As stated above, the DFT~\eqref{eq:phi-dft} contains $N_1 \times N_2$ modes $\hat{\phi}_{\vec{k}}$.
However, not all are independent, since a real-valued field has a symmetry in its Fourier transform,
\begin{equation}
    \hat{\phi}^{}_{\vec{N}-\vec{k}}
    = \hat{\phi}^{*}_{\vec{k}} \;,
\end{equation}
where the asterisk denotes complex conjugation.
We can therefore only constrain roughly half of the modes in our reconstruction, which implicitly determines the other half.
Making the cut in $k_2$, we exclude all modes with $2 k_2 > N_2$.
The remaining modes still contain the symmetry
\begin{equation}
\begin{gathered}
\hat{\phi}^{}_{N_1-k_1,0} = \hat{\phi}^{*}_{k_1,0} \;, \\
\hat{\phi}^{}_{N_1-k_1,N_2/2} = \hat{\phi}^{*}_{k_1,N_2/2}
\quad \mbox{($N_2$ even)} \;,
\end{gathered}
\end{equation}
so that we must further exclude $2 k_1 > N_1$ for $k_2 = 0$ or $2 k_2 = N_2$.

Other modes cannot be constrained because their frequency~$\vec{l}_{\vec{k}}$ vanishes, so that they do not contribute to the deflection field~\eqref{eq:alpha-dft}.
This is the case for the constant mode $(0, 0)$, which is intuitively clear, since the offset of a potential has no physical meaning.
However, we must also exclude $(N_1/2, 0)$, $(0, N_2/2)$, $(N_1/2, N_2/2)$ from the reconstruction if $N_1$, $N_2$, or both are even, which follows directly from the definition~\eqref{eq:dft-freq} of the frequencies.

Fig.~\ref{fig:dft} shows an example of the mode selection for even and odd image sizes.
The modes that can be constrained in the reconstruction form the truncated vector $\vec{\hat{\phi}}'$.
Note that all constrained modes have a symmetric partner, so our reconstruction contains exactly half of all possible values for $\vec{k}$ that do not have vanishing frequencies.

The fact that we have truncated the vector of modes~$\vec{\hat{\phi}} \to \vec{\hat{\phi}}'$ implies further changes to the reconstruction.
We cannot simply perform the estimation with a similarly truncated matrix~$\mat{P} \to \mat{P}'$, since the reconstruction requires the full $\mat{P} \vec{\hat{\phi}}$.
However, the product $\mat{P} \vec{\hat{\phi}}$ can be rewritten in terms of $\vec{\hat{\phi}}'$ and $\mat{P}'$ using the symmetry~$P_{ij,\vec{N}-\vec{k}} = P^*_{ij,\vec{k}}$,
\begin{equation}
\begin{split}
    (\mat{P} \vec{\hat{\phi}})_{ij}
    &= \sum_{\text{all $\vec{k}$}} P_{ij,\vec{k}} \, \hat{\phi}_{\vec{k}}
    = \sum_{\text{half $\vec{k}$}} \bigl\{P_{ij,\vec{k}} \, \hat{\phi}_{\vec{k}} + P_{ij,\vec{N}-\vec{k}} \, \hat{\phi}_{\vec{N} - \vec{k}}\bigr\} \\
    &= \sum_{\text{half $\vec{k}$}} \bigl\{P'_{ij,\vec{k}} \, \hat{\phi}'_{\vec{k}} + (P'_{ij,\vec{k}} \, \hat{\phi}'_{\vec{k}})^*\bigr\} \;,
\end{split}
\end{equation}
where the last equality now expresses everything in terms of the truncated quantities.
Expanding $\mat{P}'$ and $\vec{\hat{\phi}}'$ into real and imaginary parts $\mat{P}' = \mat{R} + i \, \mat{Q}$ and $\vec{\hat{\phi}}' = \vec{u} + i \, \vec{v}$, the relation is evidently still linear in $\vec{u}$ and $\vec{v}$,
\begin{equation}\label{eq:RQ-dft}
    (\mat{P} \vec{\hat{\phi}})_{ij}
    = \sum_{\text{half $\vec{k}$}} \bigl\{ 2R_{ij,\vec{k}} \, u_{\vec{k}} - 2Q_{ij,\vec{k}} \, v_{\vec{k}} \bigr\} \;,
\end{equation}

\subsection{In Spherical Coordinates}
\label{app:sphereical}

When the survey covers a significant proportion of the sky it is necessary to reconstruct the 
lensing potential on the sphere.  This requires only a change in the $\pmb{P}$ matrix first introduced in equation~\ref{eq:ave_dd}.  One approach to this problem is to expand the lensing potential in spherical harmonics instead of discrete Fourier modes as was done in section~\ref{app:DFT}.

To derive the expression for  $\pmb{P}$ we go back to the expression for the lensed flux
\begin{align}
\delta(\btheta_i) &= \delta\left(\btheta_i  - \balpha(\theta_i) \right)   \\
 &= \delta(\btheta_i ) + \balpha(\theta_i) \cdot \grad \delta(\btheta_i) \\
& = \delta_i + \balpha_i \cdot \grad \delta_i \\
&= \delta_i + \grad \phi_i \cdot \grad \delta_i
\end{align}
where $ \grad$ is the gradient with respect to angle.
To first order in the lensing potential the product of pixel values is
\begin{align}
\delta(\btheta_i) \delta(\btheta_j) = \delta_i \delta_j  + \delta_j \left[ \grad_i \phi_i \cdot \grad_i \delta_i \right] + \delta_i \left[  \grad_j \phi_j \cdot \grad_j \delta_j \right]
\end{align}

Averaging over the ansamble of $\delta(\btheta_i)$ fields gives
\begin{align}
\langle \delta(\btheta_i) \delta(\btheta_j) \rangle &= \langle \delta_i \delta_j \rangle +  \left[ \grad_i \phi_i \cdot \grad_i \langle \delta_j \delta_i \rangle \right] +  \left[  \grad_j \phi_j \cdot \grad_j \langle \delta_i\delta_j \rangle \right] \\
&= \xi_{ij} +  \left[ \grad_i \phi_i \cdot \grad_i  \xi_{ij}  \right] +  \left[  \grad_j \phi_j \cdot \grad_j   \xi_{ij}  \right]
\end{align}

The potential will be expanded in spherical harmonics
\begin{align}
\phi(\theta,\phi) = \sum_{\ell = 0}^{\infty} \sum_{m=-\ell}^{\ell} a_{\ell m} Y_\ell^m(\theta,\phi)
\end{align}
so that the gradient of potential is
\begin{align}
\grad \phi(\theta,\phi) = \sum_{\ell = 0}^{\infty} \sum_{m=-\ell}^{\ell} a_{\ell m} \pmb{\Psi}_\ell^m(\theta,\phi)
\end{align}
where $\pmb{\Psi}_\ell^m(\theta,\phi)$ are the vector spherical harmonics.

Isotropy requires that the correlation function be a function of only the angular separation between pixels, $\gamma_{ij}$, and their radial positions.  The angular separation is given by
\begin{align}
\cos( \gamma_{ij} ) = \sin(\theta_i)\sin(\theta_j) \cos\left( \phi_i - \phi_j \right) + \cos(\theta_i)\cos(\theta_j).
\end{align}
The gradient of the correlation function with respect to the position of point $i$ is
\begin{align}
\grad_i \xi_{ij} &= \grad_i \xi(\gamma_{ij},r_i,r_j) = \frac{\partial \xi_{ij}}{\partial \theta_i}  \, \hat{\btheta} + \frac{1}{\sin\theta_i} \frac{\partial \xi_{ij}}{\partial \phi_i} \, \hat{\pmb{\phi}} \\
&= \left[ \frac{\partial \gamma_{ij} }{\partial \theta_i} \hat{\btheta} + \frac{1}{\sin\theta_i} \frac{\partial \gamma_{ij}}{\partial \phi_i} \hat{\pmb{\phi}} \right]   \frac{\partial \xi(\gamma,r_i,r_j)}{\partial\gamma} \\
&= \grad_i \gamma_{ij}   \frac{\partial \xi(\gamma,r_i,r_j)}{\partial\gamma} 
\end{align}
Using this the $\pmb{P}$ matrix can be calculated as
\begin{align}
P_{ij}^{\ell m} = \left[ \grad_i \gamma_{ij} \cdot \pmb{\Psi}_\ell^m(\theta_i,\phi_i) + \grad_j \gamma_{ij} \cdot \pmb{\Psi}_\ell^m(\theta_j,\phi_j) \right]  \frac{\partial \xi(\gamma,r_i,r_j)}{\partial\gamma} 
\end{align}

We will make the approximation that pairs of points that are correlated enough to contribute significantly to the potential estimate will be separated by a small angle ($\gamma \ll \pi$).
To a good approximation we can take 
\begin{align}
 \frac{\partial \xi(\gamma,r_i,r_j)}{\partial\gamma} \simeq \bar{D}_{ij}  \frac{\partial \xi(r_\perp,r_\parallel)}{\partial r_\perp}
\end{align}
where $\bar{D}_{ij}$ is the average comoving size distance to the pixels $i$ and $j$.

%
%
%
%

\section{ \lya\ Forest Power spectrum}
\label{app:power_spectrum}
 
For the power spectrum of the \lya\ flux we use the model of \citet{2003ApJ...585...34M}
\begin{align}
P_{{\rm ly}\alpha}\left( \pmb{k} \right) = T^2(k,\mu) P_{\rm lin}(k,z) \label{eq:ly_alpha_power}
\end{align}
where $P_{\rm lin}(k,z)$ is the linear power spectrum of matter and
\begin{align}
T^2(k,\mu) = b^2 f(\mu)^2 E(k,\mu) 
\end{align}
is a transfer function with
\begin{align}
& f(\mu) =  1 + \beta \, \mu^2  \label{eq:rsd} \\
&E(k,\mu) = \exp\left[ \left(\frac{k}{k_{nl}} \right)^{a_{nl}} - \left( \frac{k}{k_p} \right)^{a_p}    - \left( \frac{k}{k_{\nu}} \mu \right)^{a_v}  \right]  \label{eq:damping}  \\
&k_{\nu} = k_{\nu_o} \left( 1 + \frac{k}{k_{\nu_p}} \right)^{a_{\nu_p} } \\
&\mu =  |\vec{k}_\parallel| / |\vec{k}|.
\end{align}
The $f(\mu)$ factor is a result of redshift distortion and $E(k,\mu)$ comes from a combination of nonlinear structure formation, thermal broadening and gas pressure effects.
 
In our simulations we use the following parameter values 
\begin{align}\nonumber
\begin{array}{ll}
 k_{nl} = 6.77 ~h \mbox{Mpc}^{-1} & k_p = 15.9 ~h  \mbox{Mpc}^{-1} \\
   k_{\nu_p} = 0.917~h  \mbox{Mpc}^{-1}  & k_{\nu_o} = 0.819 ~h  \mbox{Mpc}^{-1} \\
 \end{array}
 \end{align}
 \begin{align}\nonumber
 \begin{array}{llll}
  a_{nl} = 0.55 & a_p = 2.12 &
   a_v = 1.5 & a_{\nu_p} = 0.528 \\
   b^2 = 0.0173 & \beta = 1.58
\end{array}
\end{align}
as suggested by \citet{2003ApJ...585...34M}.

\citet{2003ApJ...585...34M} does not consider any evolution in the transfer function with redshift.  
We evolve the power spectrum only through the evolution in the linear power spectrum  
$P_{\rm lin}(k,z)$ according to the usual linear theory.  A more sophisticated model can be used
 in the future, but this seems adequate for the purposes of this paper.

\section{Calculating the \lya\ correlation function}
\label{app:lya_correlation}

\begin{figure}
 \includegraphics[width=\columnwidth]{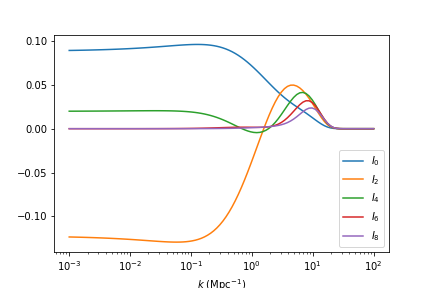}
 \caption{The factor $I_\ell(k)$ given in (\ref{eq:I_ell}) for the power spectrum discussed in section~\ref{app:power_spectrum} with $L_{\rm pix} = 1$~Mpc.}
 \label{fig:Iellofk}
\end{figure}

\begin{figure}
 \includegraphics[width=\columnwidth]{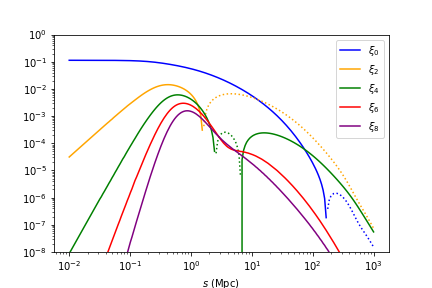}
 \caption{The first five multipoles of the correlation function (\ref{eq:xi}) as a function of separation for the power spectrum discussed in section~\ref{app:power_spectrum} with $L_{\rm pix} = 1$~Mpc.  The dotted curves are where the functions are negative.}
 \label{fig:xi_l}
\end{figure}

We need to find the correlation function for pixels (or voxels)  that are very narrow in the directions perpendicular to the line-of-sight but finite in length along the line-of-sight.  The variation in the relative flux within a pixel of comoving length $L_{\rm pix}$, or a frequency interval translated into a comoving length, will be
\begin{align}
\bar\delta(\vec{x}) &= \frac{1}{L_{\rm pix}} \int^{L_{\rm pix}/2}_{-L_{\rm pix}/2} dx' \delta( \vec{x} - \vec{x}') \\
&= \int \frac{d^3k}{(2\pi)^3} \delta_{\vec{k}} j_o\left( \frac{L_{\rm pix} k_\parallel}{2} \right) e^{i\, \vec{x} \cdot \vec{k} }
\end{align}
where $j_\ell(x)$ are the spherical Bessel functions and $\delta(\vec{x})$ is the relative density fluctuation, not the delta function.  The correlation function is
 \begin{align}
\xi(\vec{s}) &\equiv \left\langle  \bar{\delta}(\vec{x}) \bar{\delta}(\vec{x} + \vec{s} ) \right\rangle= \int \frac{d^3k}{(2\pi)^3} ~ j_o\left( \frac{L_{\rm pix} k_\parallel}{2} \right) ^2 P(\vec{k})\,  e^{i\, \vec{s} \cdot \vec{k} } \\
&= \int \frac{d k_\parallel d k_\perp}{(2\pi)^2} ~ k_\perp j_o\left( \frac{L_{\rm pix} k_\parallel}{2} \right) ^2 
J_0 \left( s_\perp k_\perp \right) P(k_\perp,k_\parallel)\,  e^{i\, s_\parallel k_\parallel } \label{eq:D4} 
 \end{align}
where $J_0(x)$ is the Bessel function of the first kind.

For calculating the correlation function it is useful to expand it in Legendre polynomials \citep{1993ApJ...417...19H}
 \begin{align}
\xi( s,\alpha ) = \sum_\ell \xi_\ell(s) P_\ell ( \cos(\alpha) )
 \end{align}
where 
$\mu = s_\parallel / |\vec{s}| = \cos(\alpha)$ 
and
 \begin{align}
\xi_\ell( s ) &= \left( \frac{2\ell +1}{2} \right)  \int_0^\pi d\alpha \,\sin(\alpha) P_\ell ( \cos(\alpha) ) \xi(s,\alpha) \\
&= \left( \frac{2\ell +1}{2} \right)  \int_{-1}^1 d\mu \, P_\ell ( \mu ) \xi(s,\mu) 
 \end{align}
Symmetry requires that $\xi_\ell( s )=0$ for odd $\ell$.
Inserting (\ref{eq:D4}) and using the integral
 \begin{align}
\int_0^\pi d\theta\sin(\theta) P_\ell\left(\cos(\theta)\right) J_o(s k_\perp \sin(\theta) ) \exp\left( i s k_\parallel \cos(\theta) \right) \nonumber \\
= 2 i^\ell P_\ell \left( cos (\alpha ) \right) j_\ell(sk)
 \end{align}
\citep{0305-4470-39-18-L06,2007JPhA...4014029C} gives
 \begin{align}
\begin{split}
\xi_\ell( s )  = & ~(-1)^{\ell/2} \frac{2\ell +1}{(2\pi)^2} \int_0^\infty dk~k^2 j_\ell ( s k) \\
& \times ~ \int_{-1}^1 d\mu~  j_o\left( \frac{L_{\rm pix} k}{2} \mu \right) ^2 P_\ell(\mu) P(k,\mu).
~~~~\ell \mbox{ even}
\end{split}
\end{align}
Substituting the \lya\ power spectrum (\ref{eq:ly_alpha_power}) gives
\begin{align} \label{eq:xi}
\xi_\ell( s )  =  \int_0^\infty \frac{dk}{(2\pi)^2 }~k^2 j_\ell ( s k)  P_{\rm lin}(k) ~ I_\ell( k) 
\end{align}
with
\begin{align}\label{eq:I_ell}
I_\ell(k) = (-1)^{\ell/2} (2\ell +1) \int_{-1}^1 d\mu~  j_o\left( \frac{L_{\rm pix} k}{2} \mu \right) ^2  T^2(k,\mu) P_\ell(\mu)
 \end{align}
For small $k$ the prefactor, $T^2(k,\mu) $, becomes independent of $k$ ($E(k,\mu) \rightarrow 1$) so $I_\ell(k)$ will be constant for large scales.
The first few $I_\ell(k)$ are shown in figure~\ref{fig:Iellofk} for the power spectrum discussed in section~\ref{app:power_spectrum}.  The first five non zero $\xi_\ell(s)$ are plotted in figure~\ref{fig:xi_l}.

\end{document}